\documentclass[prd,onecolumn,twoside,tightenlines,11pt,preprintnumbers,
               nofootinbib,notitlepage]{revtex4}

\usepackage[english]{babel}
\usepackage{amsmath,amssymb,slashed}
\usepackage{graphicx}
\usepackage[sort&compress]{natbib}
\usepackage{xcolor}
\definecolor{red}{rgb}{1.0, 0, 0}

\allowdisplaybreaks

\newcommand{\ket}[1]{\ensuremath{| #1 \rangle}}   
\newcommand{\ev}[1]{\ensuremath{\left\langle #1 %
                     \right\rangle}} 
\newcommand{\diag}{{\rm diag}} 

\begin{document}

\preprint{FERMILAB-PUB-12-038-T}

\title{Sterile neutrinos and indirect dark matter searches in IceCube}

\author{Carlos A.\ Arg\"uelles$^{1,2}$}     \email{c.arguelles@pucp.edu.pe}
\author{Joachim Kopp$^2$}                   \email{jkopp@fnal.gov}

\affiliation{${}^1$Secci\'{o}n F\'{i}sica, Departamento de Ciencias,
  Pontificia Universidad Cat\'{o}lica del Per\'{u}, Apartado 1761, Lima, Peru}

\affiliation{${}^2$Fermilab, Theoretical Physics Department, PO Box 500,
  Batavia, IL 60510, USA}

\date{March 27, 2012}

\begin{abstract}
If light sterile neutrinos exist and mix with the active neutrino flavors,
this mixing will affect the propagation of high-energy neutrinos from dark
matter annihilation in the Sun. In particular, new Mikheyev-Smirnov-Wolfenstein
resonances can occur, leading to almost complete conversion of some active
neutrino flavors into sterile states. We demonstrate how this can weaken
IceCube limits on neutrino capture and annihilation in the Sun and how
potential future conflicts between IceCube constraints and direct detection
or collider data might be resolved by invoking sterile neutrinos. We also point
out that, if the dark matter--nucleon scattering cross section and the
allowed annihilation channels are precisely measured in direct detection
and collider experiments in the future, IceCube can be used to constrain
sterile neutrino models using neutrinos from the dark matter annihilation.
\end{abstract}

\maketitle

\section{Introduction}

The hunt for dark matter is currently at a very exciting, but also somewhat
confusing stage. Many unexpected experimental results that have been reported
over the past few years can be interpreted in terms of dark matter, but all of
them could have more mundane explanations, and moreover the dark matter
interpretations of different experimental data sets do not fit together in many
cases. For instance, the signals reported by CoGeNT~\cite{Aalseth:2010vx,
Aalseth:2011wp}, DAMA~\cite{Bernabei:2008yi, Bernabei:2010mq} and
CRESST~\cite{Angloher:2011uu} appear to be in some tension with the null
results from other direct detection experiment, in particular
CDMS~\cite{Ahmed:2009zw, Ahmed:2010wy} and XENON-100~\cite{Aprile:2011hi} (see,
however, \cite{Collar:2011kf, Collar:2011wq}). Moreover, the dark matter
parameter regions favored by CoGeNT, DAMA and CRESST do not coincide under
standard assumptions on the dark matter halo~\cite{Frandsen:2011ts, Fox:2011px, Farina:2011pw,
Schwetz:2011xm, McCabe:2011sr, Kopp:2011yr} (see, however, refs.~\cite{Hooper:2011hd,
Belli:2011kw, Kelso:2011gd, Frandsen:2011gi}).  Also, if the recently observed
anomalies in the cosmic electron and positron spectra~\cite{Adriani:2008zr,
Abdo:2009zk}) are due to dark matter annihilation or decay, this would imply
dark matter masses of order 1~TeV (see, for instance, \cite{Bergstrom:2009fa}),
whereas the CoGeNT, DAMA and CRESST hints would indicate dark matter masses of
order 10~GeV.  It is thus clear that many of the potential hints for dark
matter must have other explanations, and this illustrates that a single
experiment might never be able to unambiguously identify dark matter. Only
matching detections by several different experiments would convince the
community at large that dark matter has been observed. Fortunately, the toolbox
for dark matter search is quite large: Direct detection experiments like
CoGeNT, DAMA, CRESST, CDMS and XENON-100 search for dark matter recoils on
atomic nuclei; collider searches at the Tevatron and the LHC aim to directly
produce dark matter particles and detect them through missing energy
signatures; indirect searches look for the annihilation or decay products of
astrophysical dark matter. Among the possible messengers are electrons and
positrons, anti-protons, gamma rays, and neutrinos.

A special role is played by searches for neutrinos from dark matter
annihilation in the Sun, which are carried out by the
Super-Kamiokande~\cite{Desai:2004pq, Tanaka:2011uf} and
IceCube~\cite{Abbasi:2009uz, IceCube:2011ec} collaborations.  Even though these
searches probe the products of dark matter annihilation, the expected event
rates are usually determined by the dark matter capture rate in the Sun and
thus by the dark matter--nucleus scattering cross section.  Therefore, these
searches, even though indirect, are sensitive to the same observables as direct
detection experiments and can directly test any potential direct detection
signal (provided that dark matter can annihilate and that its annihilation
products include neutrinos).  In particular, many astrophysical uncertainties,
for instance those associated with the local dark matter density, affect the
Super-Kamiokande and IceCube searches in the same way as the direct searches,
making the comparison between those experiments quite robust with respect to
astrophysics.

On the other hand, neutrinos from dark matter annihilation in the Sun are
strongly affected by neutrino oscillation physics. In this paper, we will
investigate how the oscillations pattern of high-energy neutrinos from dark
matter annihilation in the Sun can be modified by the existence of sterile
neutrinos.  Our study is motivated by the results of the
LSND~\cite{Aguilar:2001ty} and MiniBooNE~\cite{AguilarArevalo:2010wv}
experiments, as well as the reactor antineutrino
anomaly~\cite{Mueller:2011nm,Mention:2011rk, Huber:2011wv}, all of which can be
interpreted as hints for the existence of sterile neutrinos with masses of
order 1~eV~\cite{Kopp:2011qd, Giunti:2011gz, Giunti:2011hn, Abazajian:2012ys}.
(Note, however, that even models with two sterile neutrinos cannot resolve all
tension in the global data set.) We will argue that, if sterile neutrinos
exist, neutrinos from dark matter annihilation can encounter new
Mikheyev-Smirnov-Wolfenstein (MSW) resonances when propagating out of the Sun,
and that these resonances can potentially convert a large fraction of them into
undetectable sterile states.  This can weaken constraints on dark matter
annihilation in the Sun significantly.  (The existence of new MSW resonances in
the presence of sterile neutrinos has also been investigated recently in the
context of IceCube atmospheric neutrino data~\cite{Nunokawa:2003ep,
Choubey:2007ji, Razzaque:2011ab, Halzen:2011yq, Barger:2011rc, Razzaque:2012tp}.) On the
positive side, if the dark matter--nucleon scattering cross section and the
dark matter annihilation channels are precisely determined elsewhere, for
instance in direct detection and collider experiments, IceCube can be used as a
sensitive tool for constraining sterile neutrino models.

The outline of the paper is as follows: In section~\ref{sec:nuosc}, we review
the relevant aspects of the formalism of neutrino oscillations and discuss the
effect of MSW resonances on the oscillation probabilities of high-energy
neutrinos in the Sun. We then describe in section~\ref{sec:sim} how we compute
the expected neutrino signal from dark matter annihilation in the IceCube
detector, and in section~\ref{sec:limits} we show how the existence of sterile
neutrinos modifies the dark matter constraints from IceCube. We will discuss
our results and conclude in section~\ref{sec:conclusions}.

\section{Neutrino oscillations and neutrino interactions in the Sun}
\label{sec:nuosc}

Neutrinos from dark matter annihilation in the Sun probe a very unique regime
of neutrino oscillations: They are produced in a region of very high matter
density ($\sim 150$~g/cm$^3$) at the center of the Sun~\cite{Bahcall:2004pz}, but
with energies that can be much higher than those at which neutrino oscillations
in the Sun are usually studied.

In the standard three-flavor oscillation framework, it is well known from the
study of low-energy ($\mathcal{O}(\text{MeV})$) solar neutrinos that strong
transitions between electron neutrinos, $\nu_e$, and muon/tau neutrinos,
$\nu_\mu$, $\nu_\tau$, take place in a region where the number density of
electrons $N_e$ reaches a critical value, given by the
Mikheyev-Smirnov-Wolfenstein (MSW) resonance condition~\cite{Mikheyev:1986gs,
Mikheyev:1986wj, Wolfenstein:1977ue, Akhmedov:1999uz}
\begin{align}
  N_{e}^{\rm low} = a_{\rm CP} \cos\theta_{12} \frac{\Delta m_{21}^2}{2 E_\nu} \,
                    \frac{1}{\sqrt{2} G_F} \,.
  \label{eq:MSW-low}
\end{align}
Here, $E_\nu$ is the neutrino energy, $G_F$ is the Fermi constant, $\theta_{12}$
and $\Delta m_{21}^2$ are the usual solar neutrino mixing parameters, and
$a_{\rm CP} = 1$ ($-1$) for neutrinos (antineutrinos). The MSW resonance
condition can be understood if we recall that according to the Fermi theory of
weak interactions the local matter potential due to $W$ exchange with an
electron, which is felt by electron-neutrinos but not
by muon and tau neutrinos, is given by $\sqrt{2} a_{CP} G_F n_e(r)$, with
$n_e(r) = \ev{\bar{e} \gamma^0 e}$ the electron number density at a distance
$r$ from the center of the Sun. At high matter density near the center, the
flavor-diagonal MSW potential is larger than the flavor-off-diagonal neutrino
mass term $\Delta m_{21}^2 / 2E_\nu$ for multi-MeV neutrinos.  Thus
mixing between $\nu_e$ and $\nu_\mu$, $\nu_\tau$ is suppressed, and mass and
flavor eigenstate almost coincide. For instance, the flavor eigenstate
$\nu_e$ is almost equal to the mass eigenstate $\nu_2$ for
multi-MeV neutrinos produced at the center of the Sun. At low
matter density in the outer layers of the Sun, on the other hand, the mass
terms dominate over the potential term, so that the effective mixing matrix is
close to the vacuum mixing matrix, according to which $\nu_e$ is mostly
composed of $\nu_1$.  If the change in the matter density is not too fast,
neutrinos cannot ``jump'' from one mass eigenstate to another, so that a
neutrino produced as an almost pure $\nu_2$ will still be in an almost pure
$\nu_2$ state when it exits the Sun. However, its flavor composition has
changed dramatically, and in fact, the $\nu_e$ admixture to $\nu_2$ in vacuum
is given $|U_{e2}|^2 \simeq \sin^2\theta_{12} \simeq 0.31$ (using the standard
parameterization~\cite{Akhmedov:1999uz} and the current best fit
values~\cite{Schwetz:2011zk, Schwetz:2011qt} for the leptonic mixing matrix).
Thus, almost 70\% of the neutrinos are converted to $\nu_\mu$, $\nu_\tau$ on
their way out of the Sun. The flavor-conversion happens predominantly at the
transition between the matter potential-dominated and the mass mixing-dominated
regime, where the two terms are of similar magnitude. This requirement leads
precisely to the condition~\eqref{eq:MSW-low}.

For energies above $\sim 100$~MeV (not accessible with conventional solar
neutrinos), a second MSW resonance appears at a higher density
\begin{align}
  N_{e}^{\rm high} = a_{\rm CP} \cos\theta_{13} \frac{\Delta m_{31}^2}{2 E_\nu} \,
                     \frac{1}{\sqrt{2} G_F} \,.
  \label{eq:MSW-high}
\end{align}
This second resonance leads to strong $\nu_e \leftrightarrow \nu_\mu,
\nu_\tau$ transitions if $\Delta m_{31}^2 > 0$, and to strong $\bar\nu_e
\leftrightarrow \bar\nu_\mu, \bar\nu_\tau$ transitions for $\Delta m_{31}^2 < 0$.

The requirement that the change in matter density be not too fast (see above)
can be made more precise. One can show that resonant flavor transitions in
the $(ij)$-sector cease when the \emph{adiabaticity condition}~\cite{Akhmedov:1999uz}
\begin{align}
  \gamma_r \equiv \bigg( \frac{\Delta m_{ij}^2}{2 E_\nu} \sin 2\theta_{ij} \bigg)^2
                         \frac{1}{|\dot{V}|_{\rm res}} \gg 1
  \label{eq:adiabaticity}
\end{align}
is no longer fulfilled. Here, $\gamma_r$ is called the adiabaticity parameter
at the resonance and $|\dot{V}|_{\rm res}$ denotes the gradient of the MSW
potential $V = \sqrt{2} G_F N_e$ at the location of the resonance. Loss of
adiabaticity thus occurs for small mixing angles, small $\Delta m^2$ and high
energies. In the case of the resonance in the (12)-sector, which turned out to
be the solution to the long-standing solar neutrino problem, we expect
adiabatic transitions below $\sim 10$~GeV, and non-adiabatic
behavior above. (Note that, if flavor transitions of solar neutrinos were
non-adiabatic, an initial $\nu_e$ would leave the Sun not as a $\nu_2$ mass
eigenstate, but as a superposition of the form $U_{e1}^* \ket{\nu_1} + U_{e2}^*
e^{i\phi} \ket{\nu_2}$, with the oscillation phase $\phi$. After averaging over
$\phi$, this would lead to a $\nu_e$ survival probability given by $1 -
\tfrac{1}{2} \sin^2 2\theta_{12}$, in conflict with the experimental data on
solar neutrinos.)

\begin{figure}
  \begin{center}
    \begin{tabular}{c@{}c@{}c}
      \includegraphics[width=5.5cm]{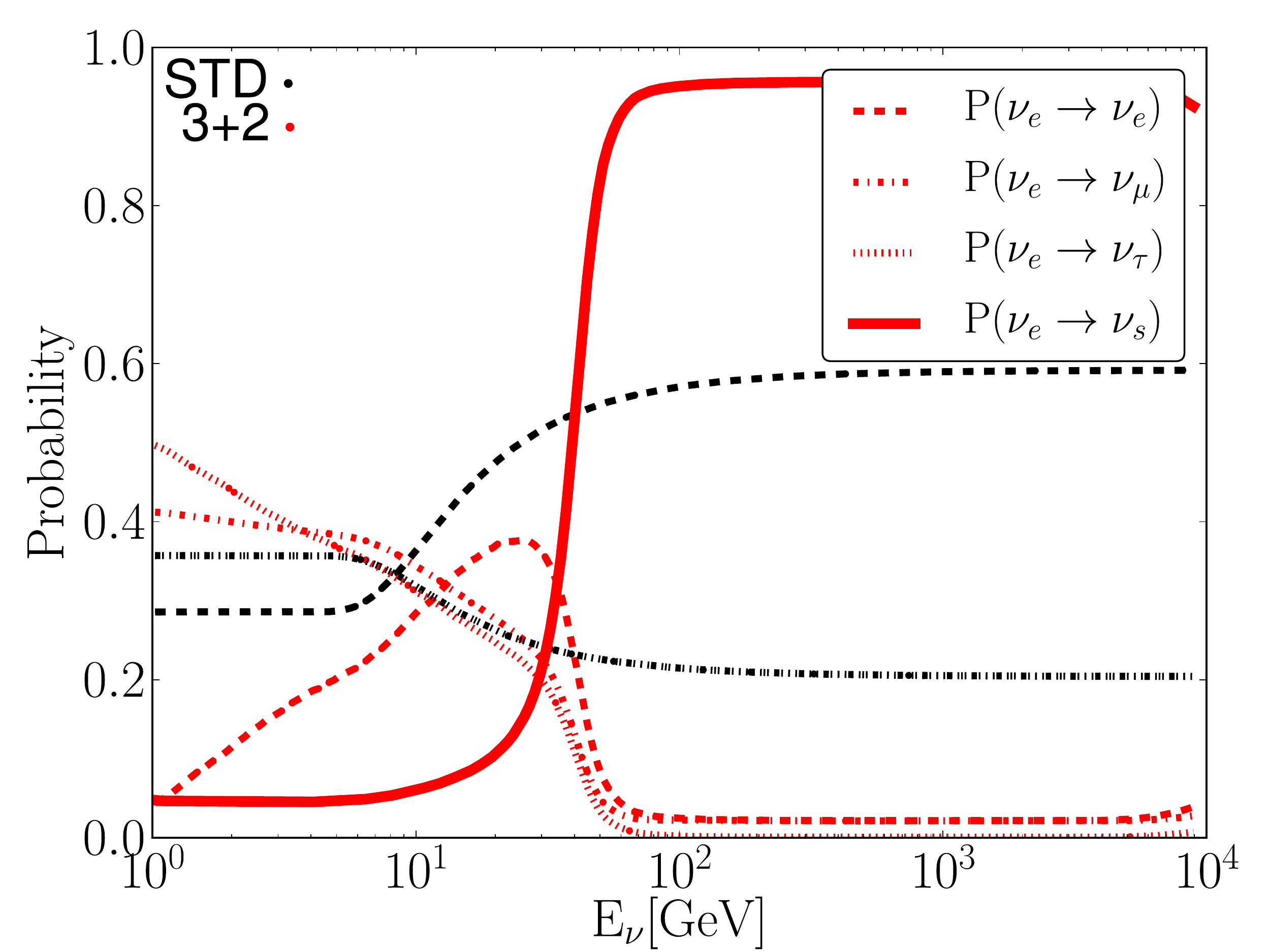} &
      \includegraphics[width=5.5cm]{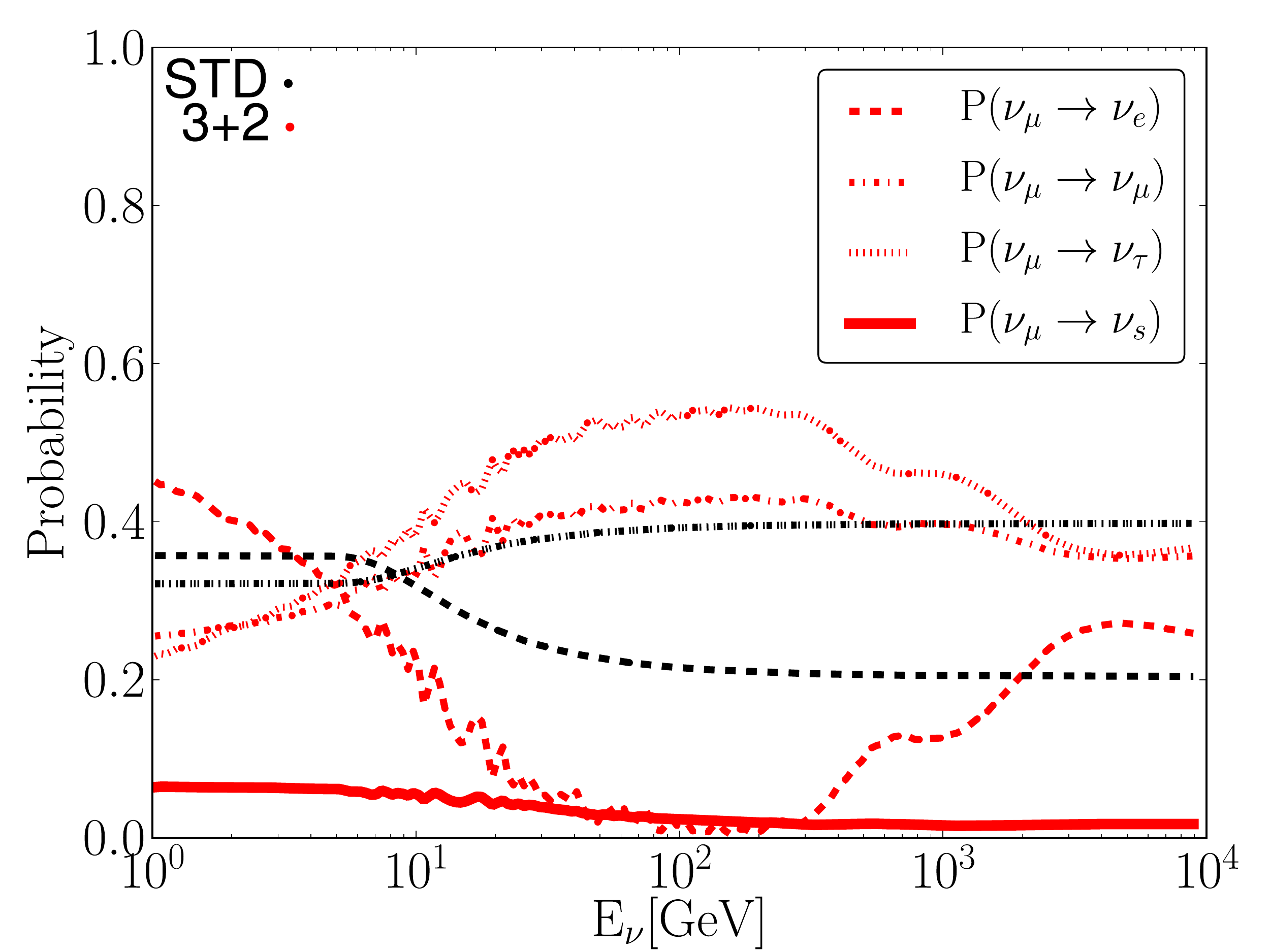} &
      \includegraphics[width=5.5cm]{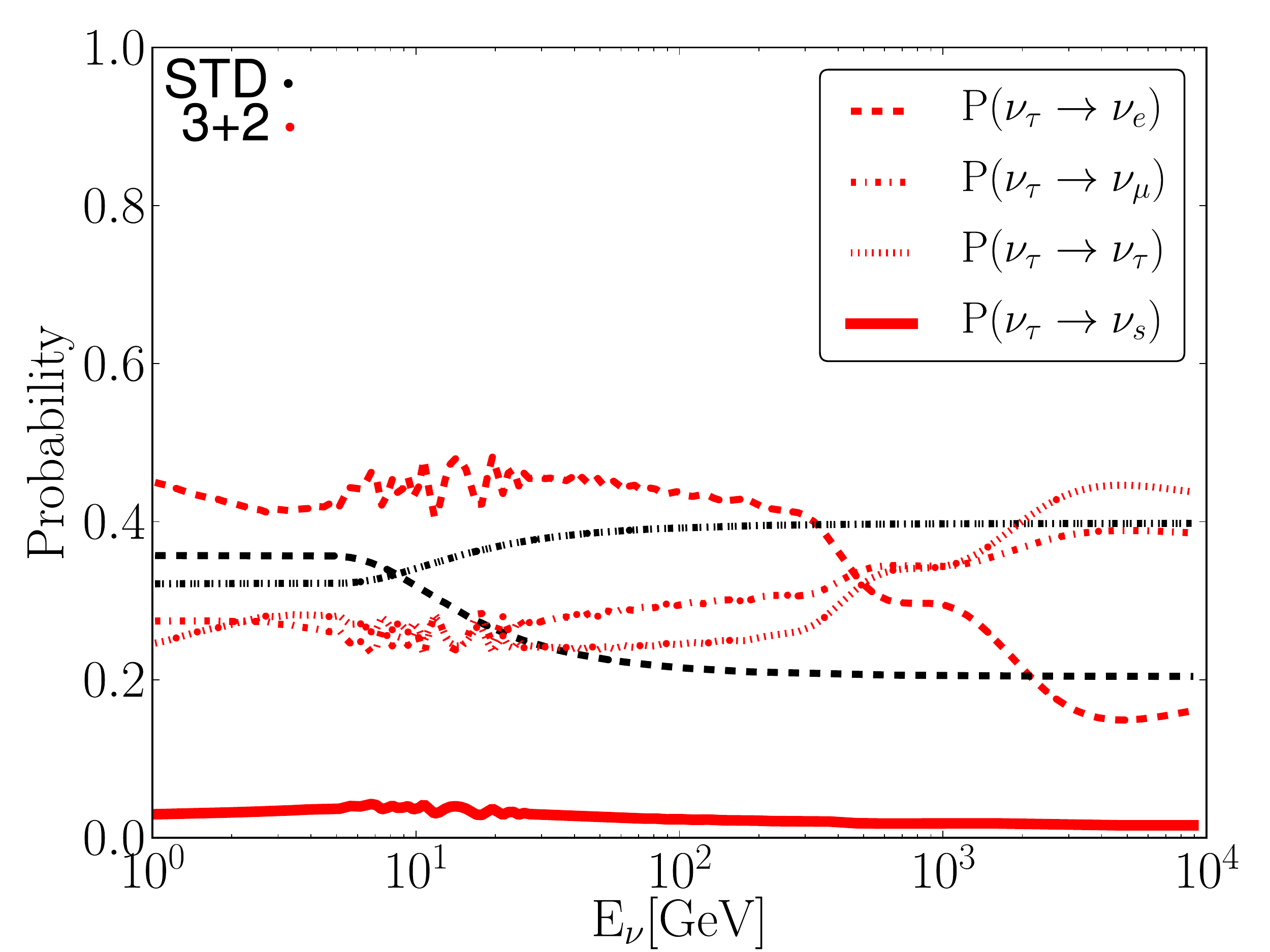} \\
      \includegraphics[width=5.5cm]{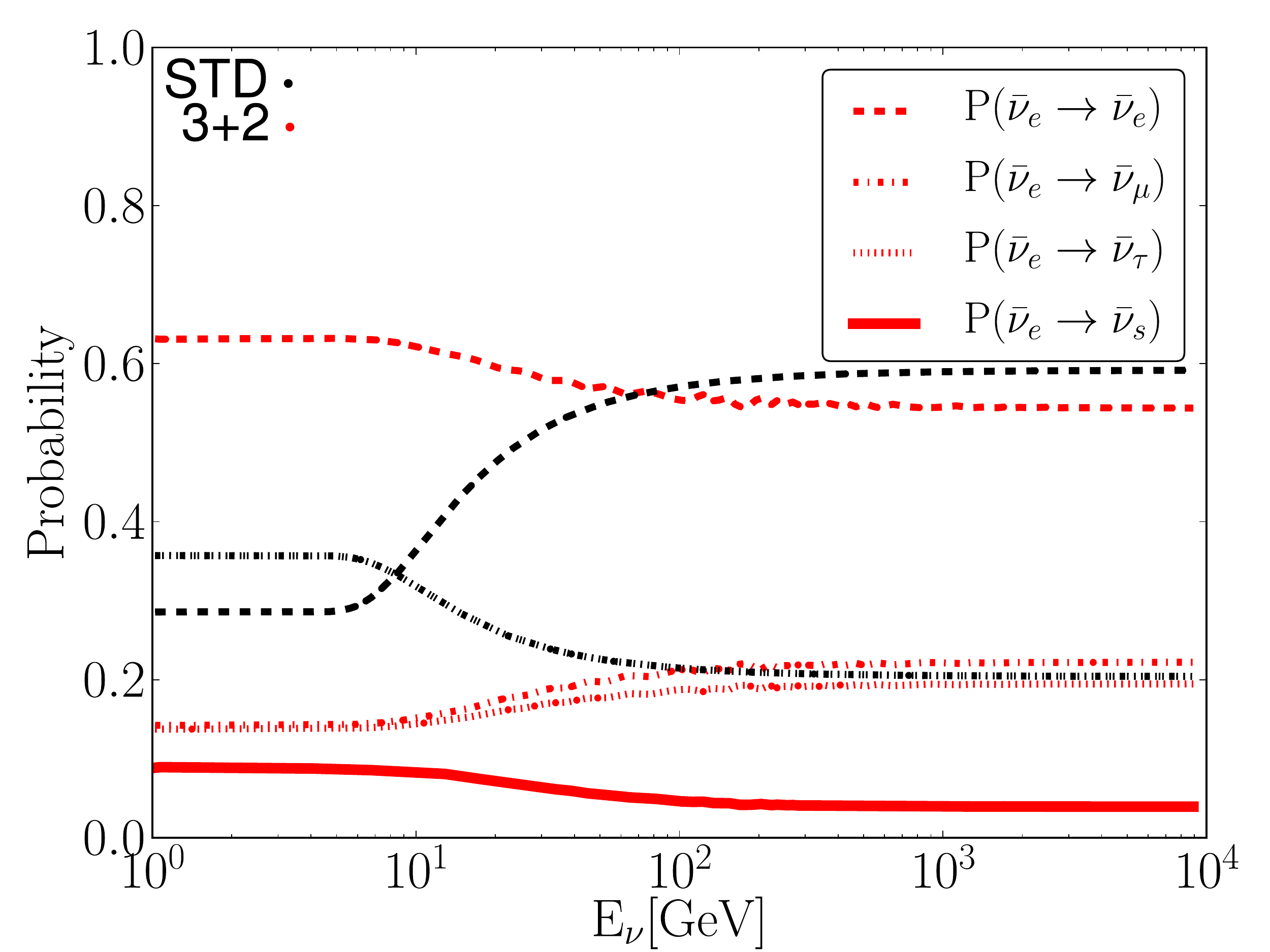} &
      \includegraphics[width=5.5cm]{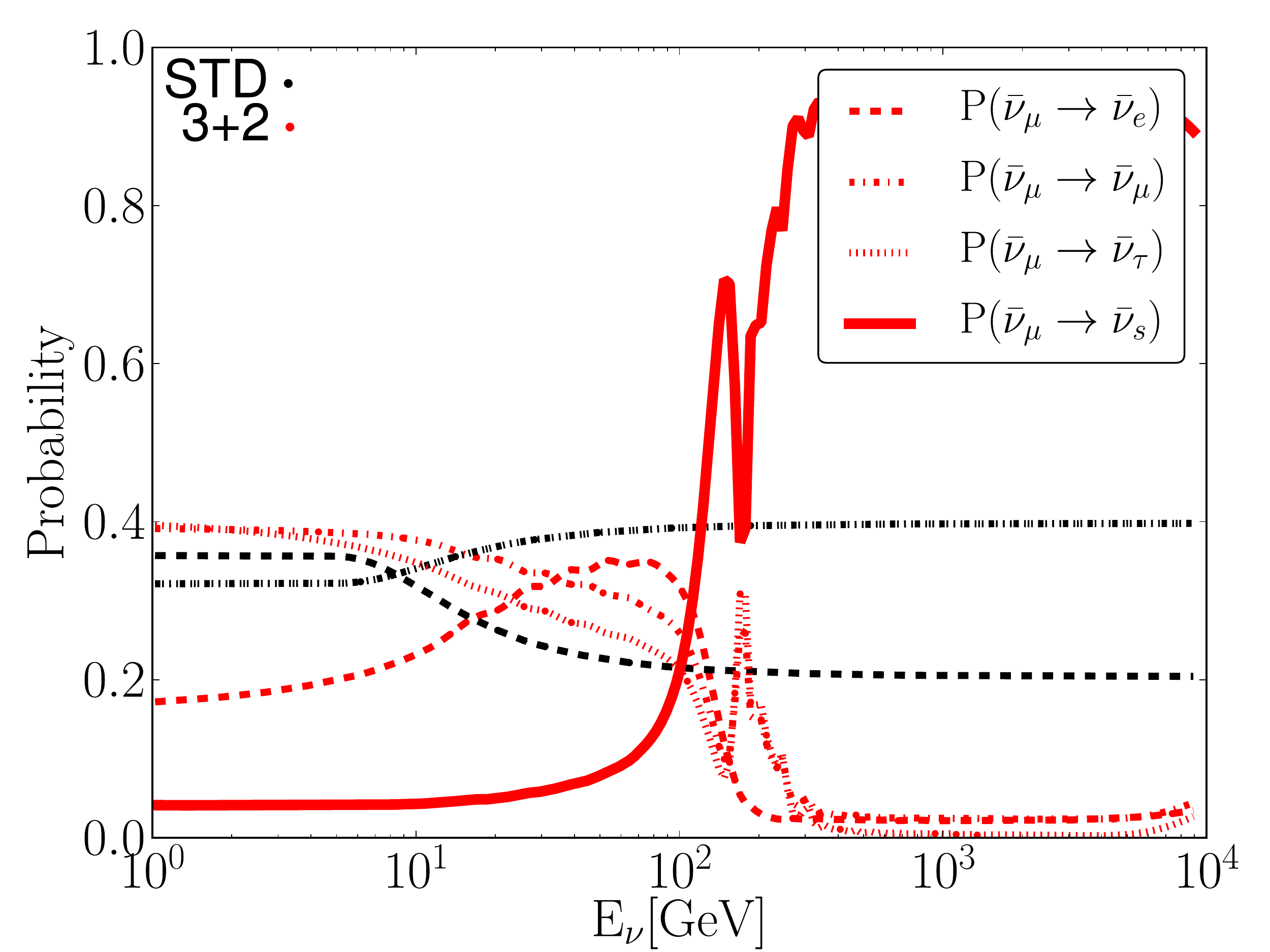} &
      \includegraphics[width=5.5cm]{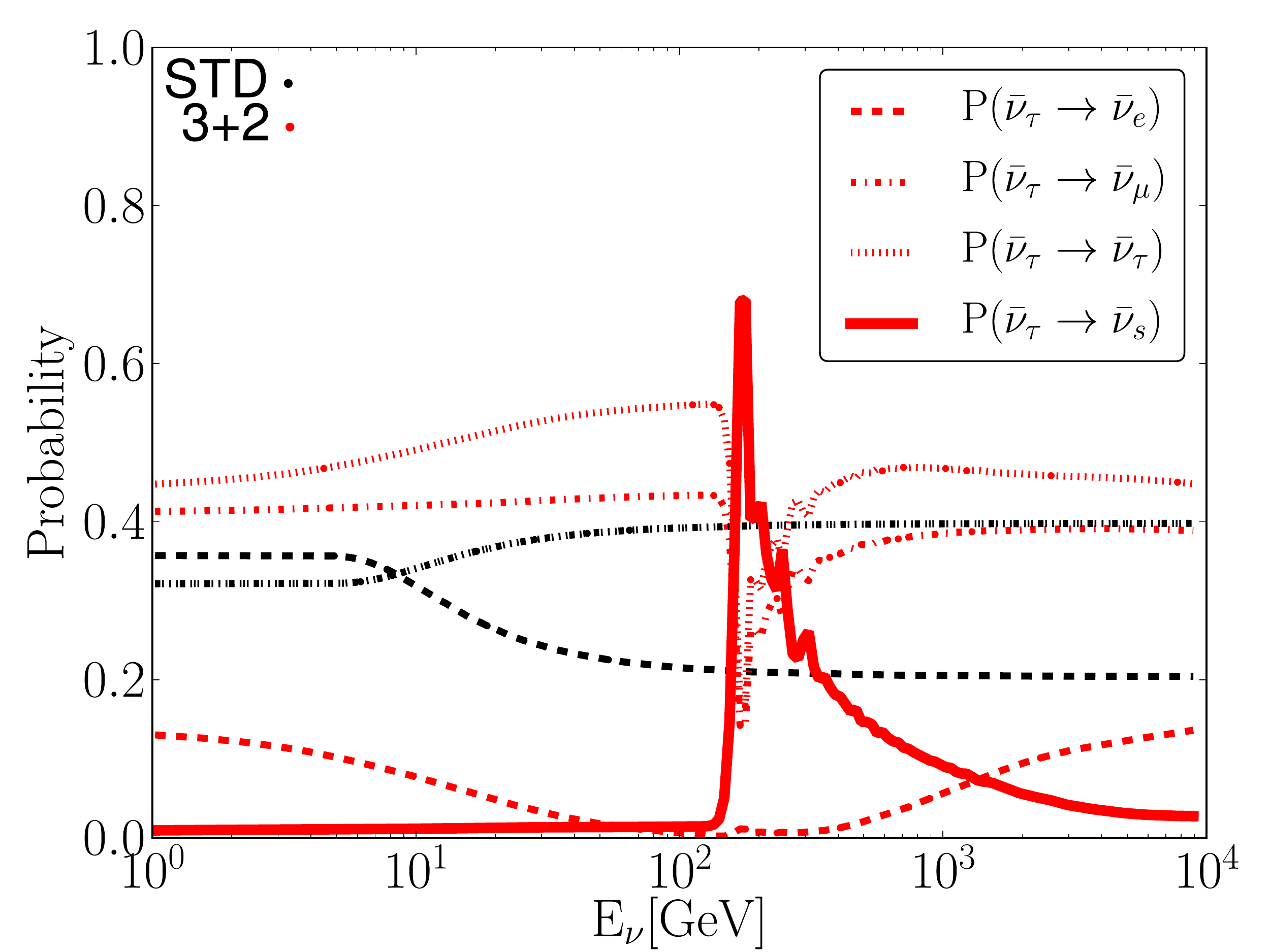} \\
    \end{tabular}
  \end{center}
  \caption{Flavor transition probabilities in the Sun as a function of energy
    for an initial $\nu_e$ (left), an initial $\nu_\mu$ (center), and an initial
    $\nu_\tau$ (right). We treat the final neutrino flux as a fully incoherent
    mixture of neutrino mass eigenstates.  The top plots are for neutrinos, the
    ones at the bottom are for anti-neutrinos. Black lines are for standard
    three-flavor oscillation, whereas red lines are for a representative
    ``$3+2$'' model with two sterile neutrinos (see text for details).
    Absorption and $\tau$ regeneration effects are neglected in these plots.
    Note that the black dotted lines ($\nu_x \to \nu_\tau$ in the SM) and the
    black dot-dashed lines ($\nu_x \to \nu_\mu$ in the SM) lie on top of each
    other since $\nu_\mu$--$\nu_\tau$ mixing is assumed to be maximal.}
  \label{fig:prob-3+2}
\end{figure}

\begin{figure}
  \begin{center}
    \begin{tabular}{c@{}c@{}c}
      \includegraphics[width=5.5cm]{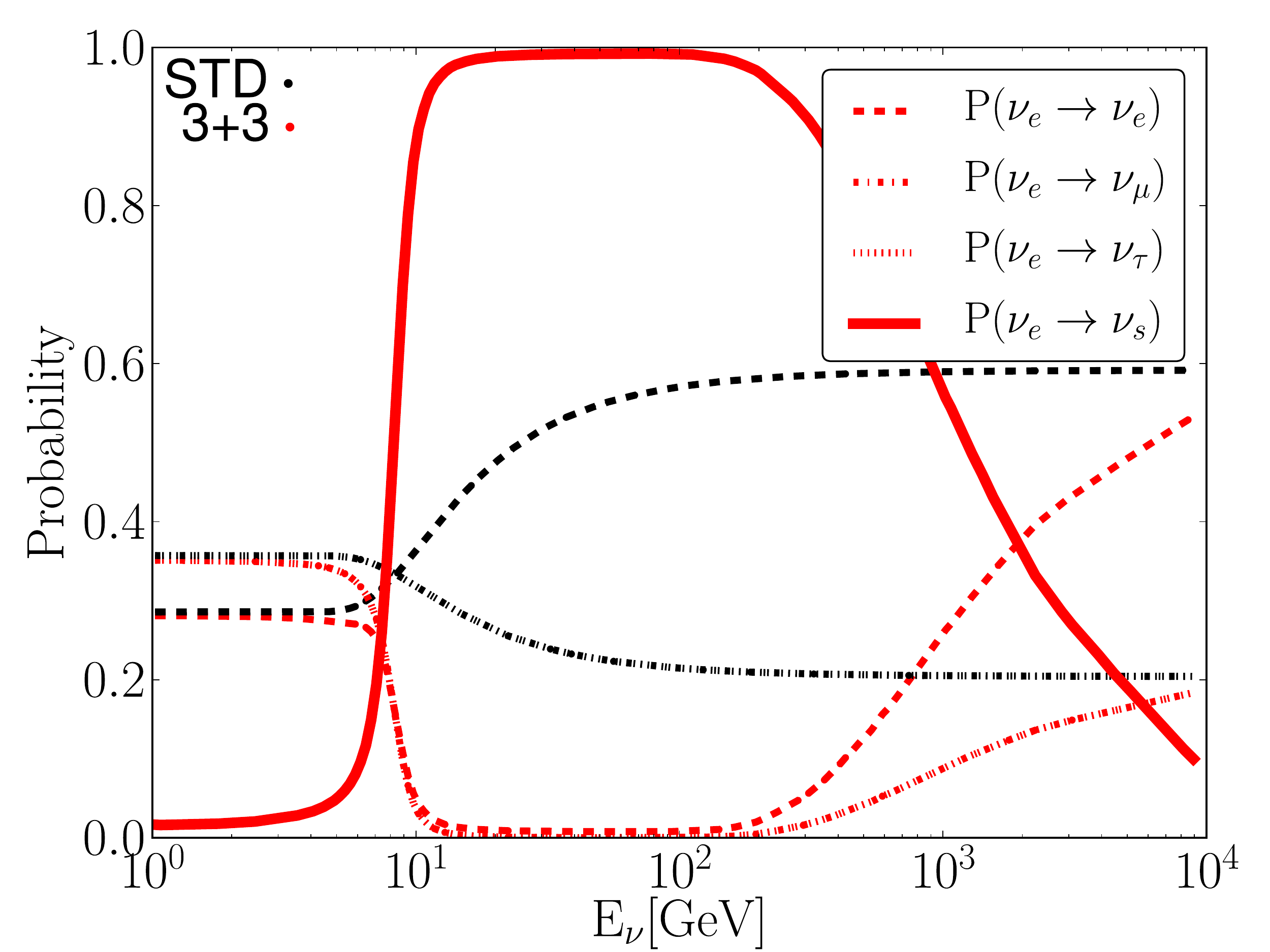} &
      \includegraphics[width=5.5cm]{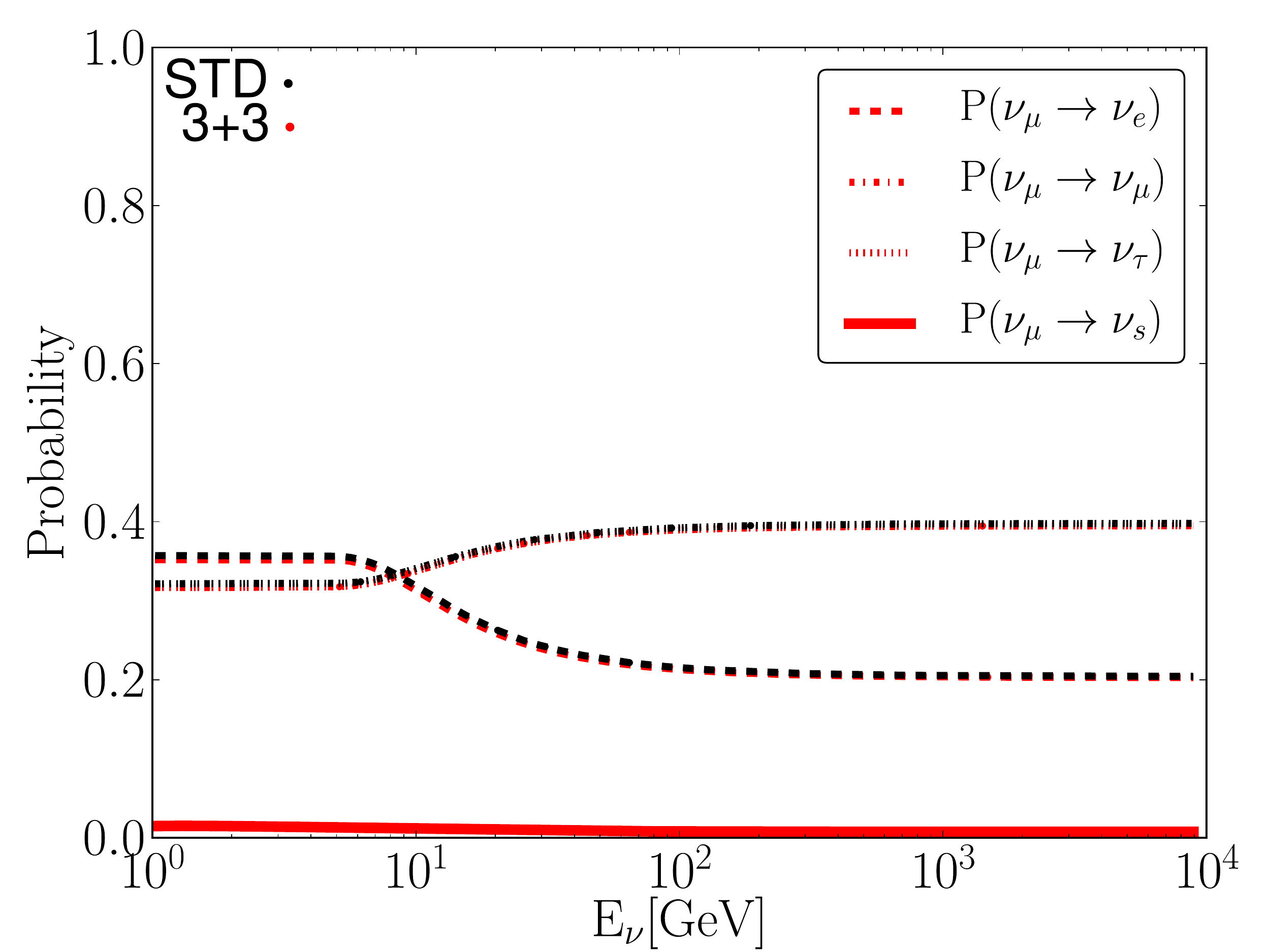} &
      \includegraphics[width=5.5cm]{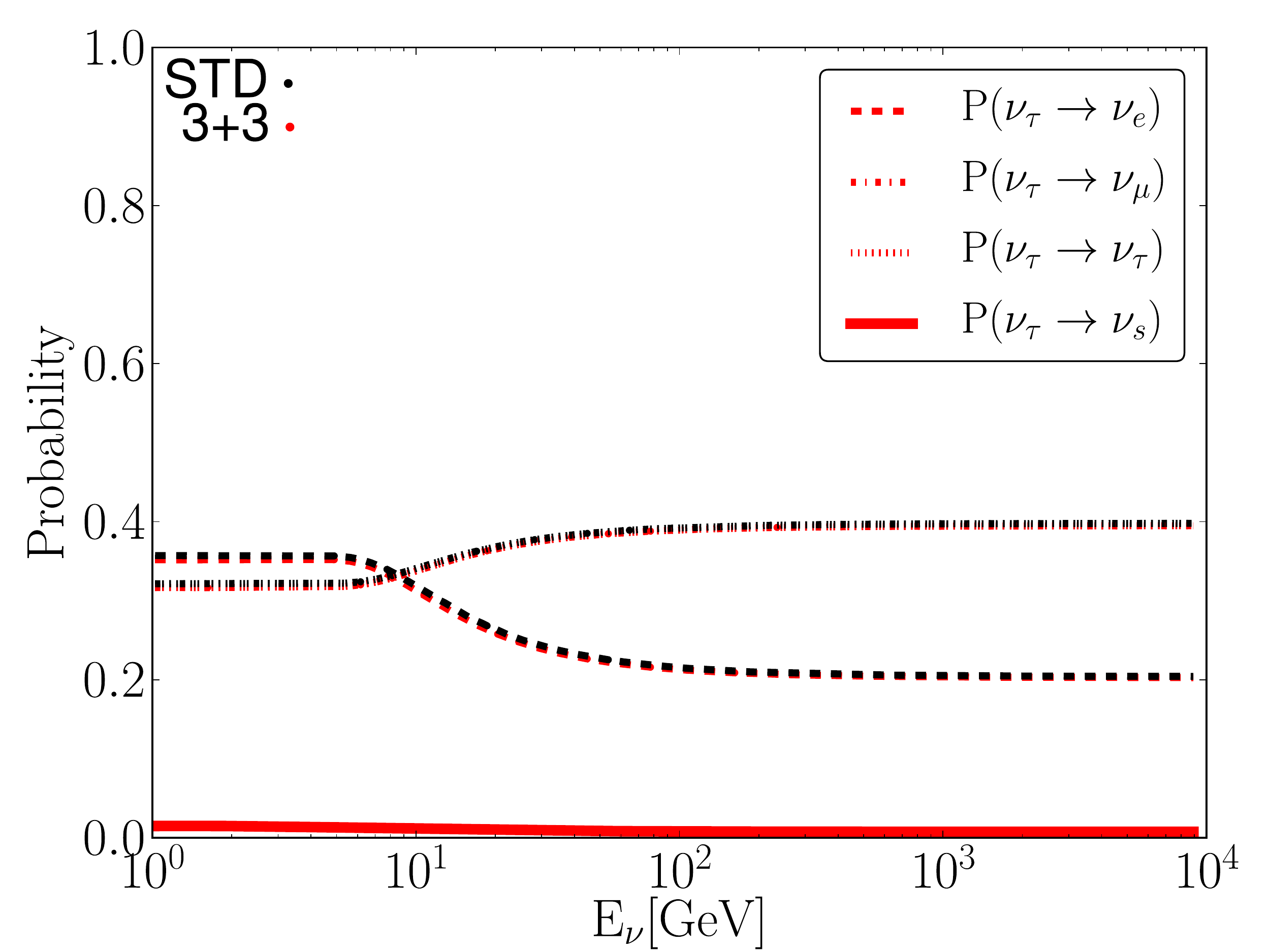} \\
      \includegraphics[width=5.5cm]{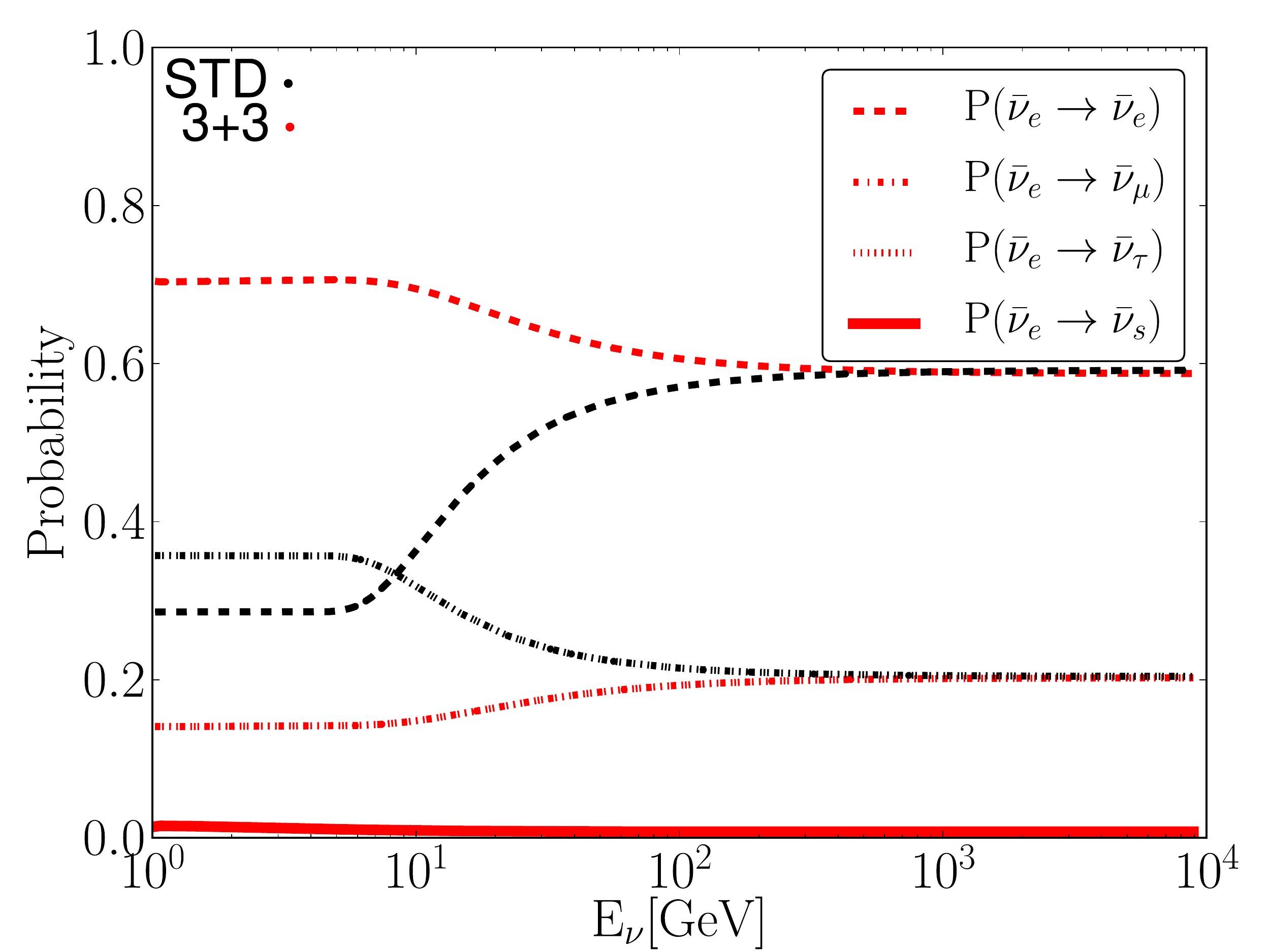} &
      \includegraphics[width=5.5cm]{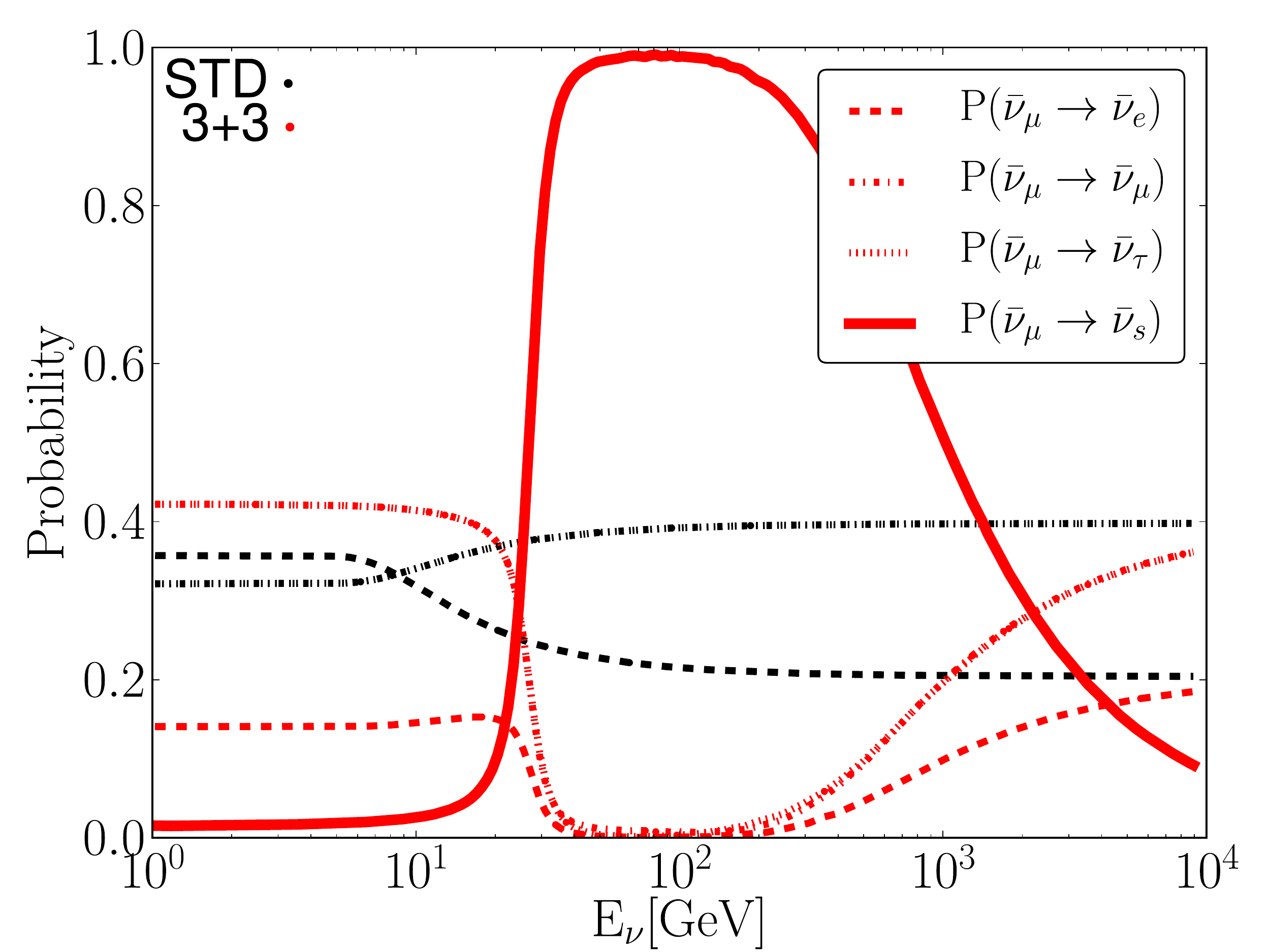} &
      \includegraphics[width=5.5cm]{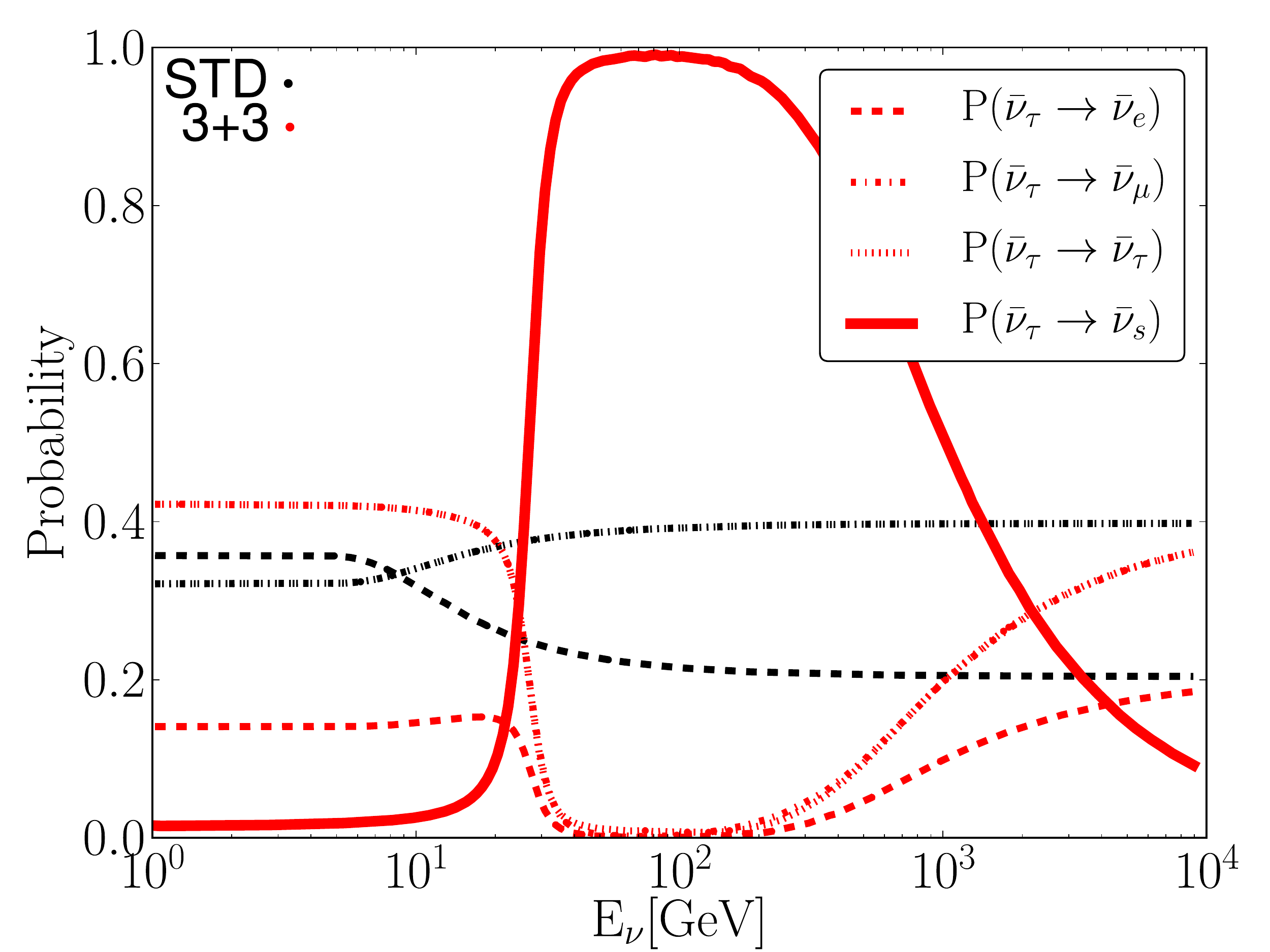} \\
    \end{tabular}
  \end{center}
  \caption{Flavor transition probabilities in the Sun as a function of energy
    for an initial $\nu_e$ (left), an initial $\nu_\mu$ (center), and an initial
    $\nu_\tau$ (right). We treat the final neutrino flux as a fully incoherent
    mixture of neutrino mass eigenstates.  The top plots are for neutrinos, the
    ones at the bottom are for anti-neutrinos. Black lines are for standard
    three-flavor oscillation, whereas red lines are for a ``$3+3$'' toy model
    with three sterile neutrinos (see text for details). Absorption and $\tau$
    regeneration effects are neglected in these plots. Note that the black
    dotted lines ($\nu_x \to \nu_\tau$ in the SM) and the black dot-dashed
    lines ($\nu_x \to \nu_\mu$ in the SM) lie on top of each other since
    $\nu_\mu$--$\nu_\tau$ mixing is assumed to be maximal.}
  \label{fig:prob-3+3}
\end{figure}

The neutrino oscillation probabilities in the Sun in the standard three-flavor
framework are plotted as a function of energy in figures~\ref{fig:prob-3+2} and
\ref{fig:prob-3+3} (black lines). Note that in computing these oscillations
probabilities, we treat the final neutrino flux as a completely incoherent
mixture of mass eigenstates. This reflects the averaging of the oscillation
probability over the size of the neutrino production region at the core of the
Sun, the annual variation in the Earth--Sun distance and the experimental
energy resolution.  The transition between the adiabatic and non-adiabatic
regimes is clearly visible in figures~\ref{fig:prob-3+2} and \ref{fig:prob-3+3}
at energies around 10~GeV. At typical solar neutrino energies of few MeV, the
$\nu_e$ survival probability has the expected value of $\sin^2 \theta_{12}
\simeq 0.3$, while in the non-adiabatic regime, it is $1 - \tfrac{1}{2} \sin^2
2\theta_{12} \simeq 0.6$.  (Small deviations from these values can arise from
the inclusion of three-flavor effects, in particular a non-zero $\theta_{13}$.)


If sterile neutrinos exist, the oscillation phenomenology becomes much richer.
Even if \emph{vacuum} oscillations between active and sterile neutrino flavors
are negligible because of small mixing angles, active--sterile oscillations
\emph{in matter} can be significant, in particular at the high energies
relevant to neutrinos from dark matter annihilation.  The $n$-flavor MSW
potential has the form 
\begin{align}
  V = (n_e - n_n/2, -n_n/2, -n_n/2, 0, \dots) \,,
  \label{eq:MSW}
\end{align}
where the terms containing the neutron density $n_n$ originate from coherent forward
scattering through $Z^0$ exchange. These terms are usually neglected in the
three-flavor framework since they are flavor-universal and therefore cannot
contribute to oscillations among active neutrinos. However, they become
relevant in the presence of sterile states. In particular, there will be
additional MSW resonances whenever any of the matter potential terms becomes
equal to any of the mass terms in the Hamiltonian.  These MSW resonance can
lead to nearly complete conversion of certain neutrino (or antineutrino)
flavors into sterile states on the way out of the Sun.

To illustrate this observation, which is the main topic of this paper, we
consider a sterile neutrino scenario similar to the one that has been shown in
Refs.~\cite{Kopp:2011qd, Abazajian:2012ys} to provide a reasonably good fit to
the global neutrino data, including the anomalous LSND and MiniBooNE results.
The model has two sterile neutrino flavors $\nu_{s1}$, $\nu_{s2}$ and two new
mass eigenstates $\nu_4$, $\nu_5$ with mixing parameters
\begin{align}
  \begin{gathered}
    \sin^2 \theta_{12} = 0.32  \qquad
    \sin^2 2\theta_{13} = 3 \times 10^{-3}  \qquad
    \sin^2 \theta_{23} = 0.45   \\
    \Delta m_{21}^2 = 7.6 \times 10^{-5} \ \text{eV}^2 \qquad
    \Delta m_{31}^2 = 2.38 \times 10^{-3} \ \text{eV}^2 \\
    \Delta m_{41}^2 = 0.47 \ \text{eV}^2 \qquad
    \Delta m_{51}^2 = 0.90 \ \text{eV}^2 \\
    \sin^2 2\theta_{14} = 0.060 \qquad \\
    \sin^2 2\theta_{24} = 0.055 \qquad
    \sin^2 2\theta_{34} = 0.000 \qquad 
    \sin^2 2\theta_{15} = 0.086 \qquad \\
    \sin^2 2\theta_{25} = 0.088 \qquad 
    \sin^2 2\theta_{35} = 0.002 \qquad
    \delta_{13} = 1.47 \pi \qquad
    \delta_{14} = 0.77 \pi \qquad
    \delta_{15} = 1.086 \pi \qquad
  \end{gathered}
  \label{eq:3+2-params}
\end{align}
Here, we use the parameterization
\begin{align}
  U_{3+2} = R_{45} R_{35} R_{25} R_{15}^\delta R_{34} R_{24} R_{14}^\delta R_{23} R_{13}^\delta R_{12}
\end{align}
for the leptonic mixing matrix,
where $R_{ij}$ denotes a rotation matrix in the $(ij)$ plane with rotation
angle $\theta_{ij}$, and $R_{ij}^\delta$ denotes a similar rotation matrix
which in addition carries a complex phase $\delta_{ij}$:
\begin{align}
  R_{ij} = \begin{pmatrix}
              \ddots & \\
                     & \cos\theta_{ij}  & \cdots & \sin\theta_{ij} \\
                     & \vdots           &        & \vdots \\
                     & -\sin\theta_{ij} & \cdots & \cos\theta_{ij} \\
                     &                  &        &                 & \ddots
           \end{pmatrix} \,,\qquad
  R_{ij}^\delta = \begin{pmatrix}
              \ddots & \\
                     & \cos\theta_{ij}  & \cdots & \sin\theta_{ij} e^{-i \delta_{ij}}\\
                     & \vdots           &        & \vdots \\
                     & -\sin\theta_{ij} e^{i \delta_{ij}} & \cdots & \cos\theta_{ij} \\
                     &                  &        &                 & \ddots
           \end{pmatrix} \,.
\end{align}
In a ``$3+2$'' scenario like equation~\eqref{eq:3+2-params}, the new MSW
resonances converting active neutrinos into sterile ones affect antineutrinos
more strongly than neutrinos, but since neutrino cross sections are larger than
antineutrino cross sections, we expect the impact of sterile neutrinos on dark
matter searches to be only moderate, especially in detectors like IceCube and
Super-Kamiokande which cannot distinguish neutrinos from antineutrinos.
(Below, we will also discuss a $3+3$ toy model in which effects are larger.)

The neutrino oscillation probabilities in the Sun for this sterile neutrino
scenario are shown in figure~\ref{fig:prob-3+2} as red curves.  The most
striking feature is the strong conversion of $\bar\nu_\mu$ (and to some degree
also $\bar\nu_\tau$) into sterile neutrinos at energies above $\sim 200$~GeV.
Indeed, we can see from equation~\ref{eq:MSW-low} (with the replacements
$\theta_{12} \to \theta_{14} \simeq 0$, $\Delta m_{21}^2 \to \Delta m_{41}^2
\simeq -1$~eV$^2$, and $N_e \to -N_n/2$) that above $E_\nu \sim 100$~GeV, the MSW
resonance between active and sterile neutrinos lies within the Sun. Therefore,
high energy $\bar\nu_\mu$ and $\bar\nu_\tau$ produced from dark matter
annihilation at the center of the Sun will be almost fully converted into
sterile neutrinos, leaving as detectable states only neutrinos, and
antineutrinos from the $\bar\nu_e$ component of the primary flux.  For a given
dark matter mass, annihilation channel and annihilation cross section, the
expected event number in a neutrino detector is thus reduced, so that
experimental constraints on dark matter annihilation in the Sun become weaker.

In addition to the $3+2$ scenario, we are also going to consider a $3+3$ toy
model with 3 sterile neutrinos. The oscillation parameters in this model are
chosen such that each active neutrino flavor eigenstate mixes with only one of
the sterile neutrinos. This can be achieved by choosing mass squared difference
$\Delta m_s^2 \equiv \Delta m_{41}^2 \simeq \Delta m_{52}^2 \simeq \Delta
m_{63}^2$ and mixing angles $\theta_s \equiv \theta_{14} \simeq \theta_{25} \simeq
\theta_{36}$ (all other active--sterile mixing angles are zero), so that the
sterile neutrino sector is a mirror image of the active neutrino sector as far
as vacuum oscillations are concerned. (Similar models have been considered
in~\cite{Nelson:2007yq}.)
The parameterization of the leptonic
mixing matrix is here
\begin{align}
  U_{3+3} = R_{36} R_{25} R_{14} R_{23} R_{13}^\delta R_{12}
\end{align}
Unless specified otherwise, we choose $\Delta m_{41}^2 = 0.1$~eV$^2$ and
$\sin^2 2\theta_s = 0.03$. In general, if $\Delta m_{41}^2$, $\Delta m_{52}^2$,
$\Delta m_{63}^2 \gg \Delta m_{21}^2$, $|\Delta m_{31}^2|$, conversions of
active neutrinos into sterile neutrinos can be understood in a simple
two-flavor framework as long as the distance travelled by the neutrinos is much
shorter than the active neutrino oscillation lengths $L^{\rm osc}_{21} = 4\pi E_\nu
/ \Delta m_{21}^2$ and $L^{\rm osc}_{31} = 4\pi E_\nu / |\Delta m_{31}^2|$. This
remains true even in matter. In this case, the effective two-flavor
oscillations between an active flavor and its corresponding sterile flavor are
affected by an MSW resonance. The resonance conditions are, in analogy to
equations~\eqref{eq:MSW-low} and \eqref{eq:MSW-high}:
\begin{align}
  N_e &= a_{\rm CP} \cos\theta_{14} \frac{\Delta m_{41}^2}{2 E_\nu} \,
                    \frac{1}{\sqrt{2} G_F} \,,
                    & (\text{$\nu_e \leftrightarrow \nu_{s1}$ transitions})
  \label{eq:MSW-3+3-e} \\
  -\frac{N_n}{2} &= a_{\rm CP} \cos\theta_{25} \frac{\Delta m_{52}^2}{2 E_\nu} \,
                    \frac{1}{\sqrt{2} G_F} \,,
                    & (\text{$\nu_\mu \leftrightarrow \nu_{s2}$ transitions})
  \label{eq:MSW-3+3-mu} \\
  -\frac{N_n}{2} &= a_{\rm CP} \cos\theta_{36} \frac{\Delta m_{63}^2}{2 E_\nu} \,
                    \frac{1}{\sqrt{2} G_F} \,.
                    & (\text{$\nu_\tau \leftrightarrow \nu_{s3}$ transitions})
  \label{eq:MSW-3+3-tau}
\end{align}
We see from these equations that the resonance between $\nu_e$ and the first
sterile flavor eigenstate $\nu_{s1}$ will be in the neutrino sector ($a_{\rm CP} = 1$),
whereas the $\nu_\mu \leftrightarrow \nu_{s2}$ and $\nu_\tau \leftrightarrow \nu_{s3}$
resonances affect antineutrinos ($a_{\rm CP} = -1$). This behavior is  reflected
in figure~\ref{fig:prob-3+3}, where we show the flavor transition probabilities
for all oscillation channels in the $3+3$ model as a function of energy. We see that
in a large energy range $\nu_e$, $\bar\nu_\mu$ and $\bar\nu_\tau$ are almost fully
converted into sterile states. We expect that this will lead to a considerable
weakening of the limits IceCube can set on dark matter capture and annihilation in
the Sun.

Note that this weakening could be even more pronounced if the mostly active
neutrino mass eigenstates were \emph{heavier} than the mostly sterile ones,
since in that case the MSW resonances for second and third generation neutrinos
would move from the antineutrino sector to the neutrino sector, which is more
important for IceCube's dark matter search because neutrino interaction cross
sections are about a factor of 3 larger than antineutrino cross sections. We
do not consider this possibility here since relatively heavy active neutrinos
would be in potential conflict with cosmology~\cite{Hamann:2010bk,
Giusarma:2011ex, Hamann:2011ge, Giusarma:2011zq}.  (These conflict can
potentially be avoided in non-minimal cosmologies~\cite{Hamann:2011ge,
Giusarma:2011zq} and in models where the relic abundance of sterile neutrinos
is reduced, see for instance references~\cite{Gelmini:2004ah, Smirnov:2006bu}
for a discussion of such models.)

\begin{figure}
  \begin{center}
    \begin{tabular}{c@{}c@{}c}
      \hspace*{-0.3cm}\includegraphics[width=6.0cm]{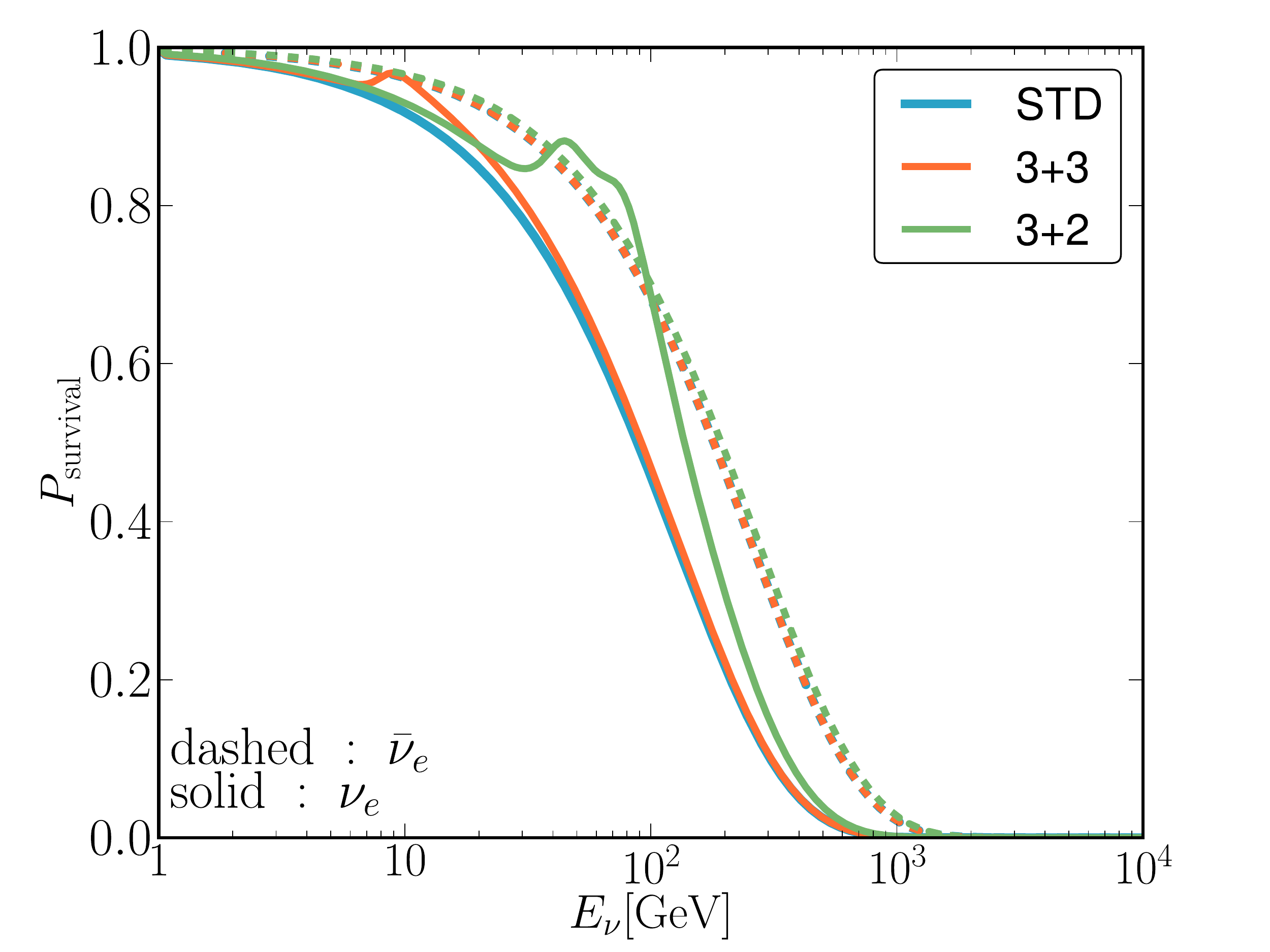}
      \hspace*{-0.5cm}\includegraphics[width=6.0cm]{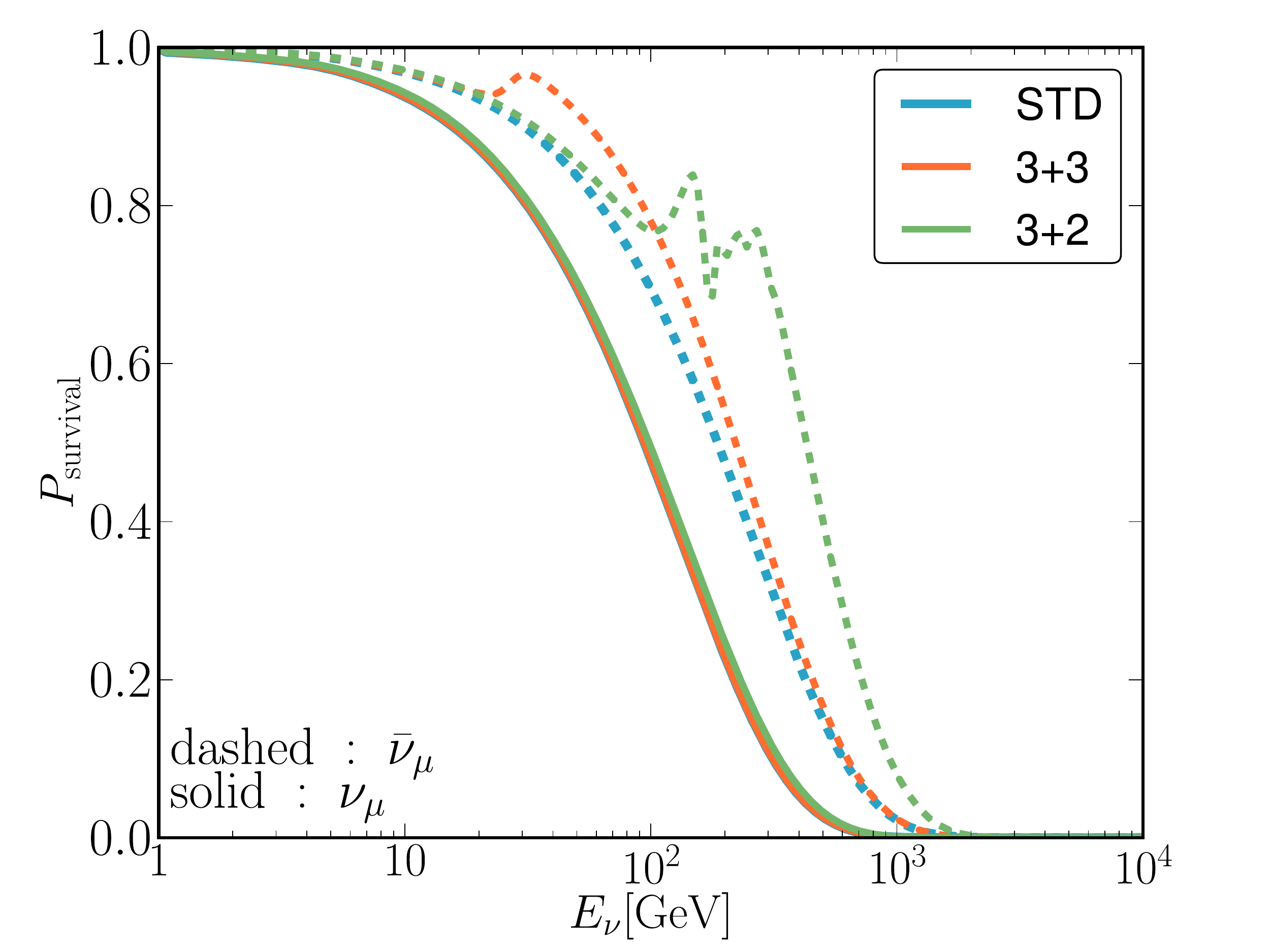}
      \hspace*{-0.5cm}\includegraphics[width=6.0cm]{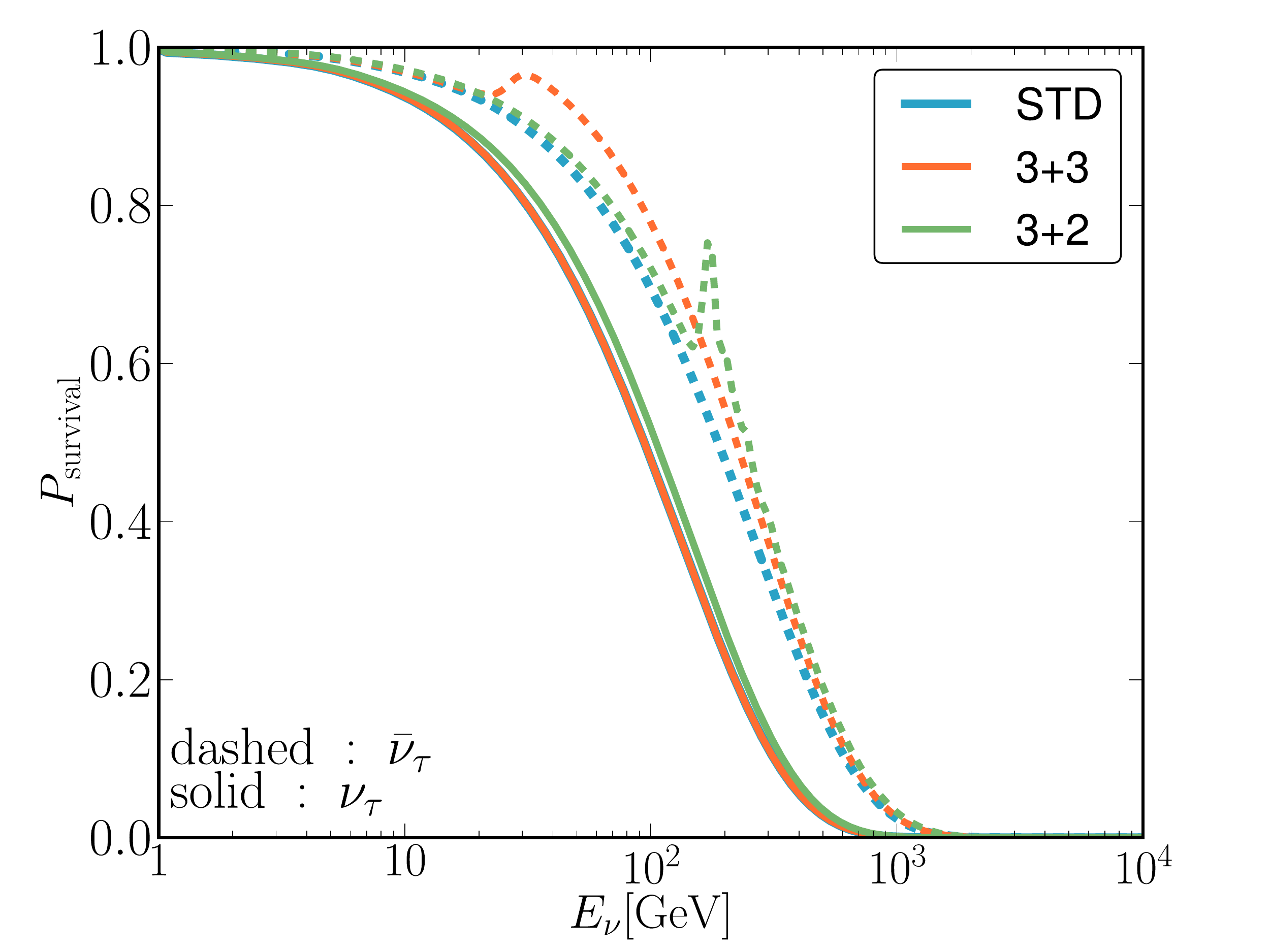}
    \end{tabular}
  \end{center}
  \caption{Survival (= non-interaction) probabilities for high-energy neutrinos
    from dark matter annihilation on their way out of the Sun. This plot shows
    only the effect of neutrino absorption or scattering, not that of flavor
    transitions. Flavor transitions do, however, affect the absorption
    probability indirectly: for instance a neutrino that has been converted
    into a sterile flavor can no longer be absorbed.  We show results here for
    standard three-flavor oscillations, for the best fitting $3+2$ model, and
    for a $3+3$ toy model. The features in the $3+2$ and $3+3$ curves are due
    to the interplay of active--sterile conversion and active neutrino
    interactions.}
  \label{fig:survival-prob}
\end{figure}

Apart from oscillation, the propagation of high-energy neutrinos through the
Sun is also affected by non-coherent neutral current (CC) and charged current
(CC) interactions. NC interactions change the neutrino energy, whereas CC
interactions lead to absorption and possible reemission of neutrinos in the
decay of secondary $\mu$ or $\tau$ leptons. Since secondary muons are usually
thermalized before they decay, reemission of \emph{high-energy} neutrinos is
only possible in the case of $\nu_\tau + X \to \tau + X'$ CC interactions
(``$\tau$ regeneration''). In figure~\ref{fig:survival-prob} we plot
the non-interaction (``survival'') probability $P_{\rm survival}$ for neutrinos
from dark matter annihilation on their way out of the Sun as a function of the
neutrino energy. $P_{\rm survival}$ can be calculated as
\begin{align}
  P_{\rm survival}(E_\nu) = \exp\bigg[-\int\!dx\,\rho(x)
       \left( \sigma_{\rm CC_e} p_e(x) + \sigma_{\rm CC_\mu} p_\mu(x)
       + \sigma_{\rm CC_\tau} p_\tau(x) + \sigma_{\rm NC} p_{\rm active}(x) \right) \bigg] \,,
\end{align}
where $p_\alpha(x)$ is the probability of the neutrino being in the flavor state $\alpha$
at position $x$, $p_{\rm active}(x) \equiv \sum_{\alpha = e,\mu,\tau} p_\alpha(x)$,
and the integral runs over the neutrino trajetory,

\section{Simulation techniques}
\label{sec:sim}

To estimate quantitatively how existing limits on dark matter annihilation in
the Sun are modified in the presence of sterile neutrinos, we have carried out
numerical simulations. We compute the dark matter capture rate as a function of
the dark matter mass and scattering cross section using the formulae
from~\cite{Gould:1992} and assuming a local WIMP density of 0.3~GeV/cm$^3$ with
an isothermal velocity distribution and velocity dispersion 220~km/sec.
We assume the annihilation cross section to be large enough for the capture and
annihilation reactions to be in equilibrium, so that the annihilation rate is
equal to half the capture rate. We use initial neutrino spectra
from~\cite{WimpSimHP}, which were generated using
{\tt WimpSim}~\cite{Blennow:2007tw}.

To propagate the neutrinos out of the Sun, we use our own Monte Carlo code,
which is capable of working with an arbitrary number of neutrino flavors $n$,
and simulates $n$-flavor oscillations in matter as well as NC and CC neutrino
scattering in the Sun, including $\tau$ regeneration.  We use the {\tt nusigma}
package~\cite{Blennow:2007tw,WimpSimHP} to calculate the neutrino cross
sections, and TAUOLA~\cite{Jadach:1993hs} for decaying secondary $\tau$'s.

In practice, we proceed as follows: We propagate the $n$-component neutrino
state vector $\psi(t)$ out of the Sun using the {\tt rkf45}
Runge-Kutta-Fehlberg algorithm from the GNU Scientific
Library~\cite{Galassi:GSL} to solve the evolution equation
\begin{align}
  i \frac{d}{dt} \psi(t) = \frac{1}{2E_\nu} U
  \begin{pmatrix}
    0 &                 &                 &                 & \\
      & \Delta m_{21}^2 &                 &                 & \\
      &                 & \Delta m_{31}^2 &                 & \\
      &                 &                 & \Delta m_{41}^2 & \\
      &                 &                 &                 & \ddots
  \end{pmatrix} U^\dag \psi(t)
  + \sqrt{2} G_F
  \begin{pmatrix}
    N_e(t) - \tfrac{N_n(t)}{2} &                    &                    &     \\
                               & -\tfrac{N_n(t)}{2} &                    &     \\
                               &                    & -\tfrac{N_n(t)}{2} &     \\
                               &                    &                    & 0   \\
                               &                    &                    &   & \ddots
  \end{pmatrix} \psi(t) \,.
\end{align}
After each Runge-Kutta step, we determine randomly if the neutrino undergoes an
incoherent interactions during that step. The probability for a CC or NC
interaction is given by $P_{\rm CC/NC} = \sigma_{\rm CC/NC} / (\sigma_{\rm CC}
+ \sigma_{\rm NC}) \times \big[1 - \exp\big(-\Delta r \, n(r) \, (\sigma_{\rm
CC} + \sigma_{\rm NC})\big)\big]$, where $n(r)$ is the local nucleon number
density, $\sigma_{\rm CC (NC)}$ is the charged current (neutral current)
neutrino--nucleon scattering cross section, and $\Delta r$ is the current
Runge-Kutta step size. If it is determined that the neutrino interacts through
a neutral current, its energy after the interaction is picked randomly from the
final state energy spectrum calculated using
{\tt nusigma}~\cite{Blennow:2007tw,WimpSimHP}. Since we are treating neutrino
propagation as a one-dimensional problem, we assume that the direction of
travel does not change, and we continue to propagate the neutrino radially
outward.  In the case of a $\nu_e$ or $\nu_\mu$ charged current interactions,
we simply discard the neutrino. In a charged current $\nu_\tau$ interactions,
the original neutrino is also absorbed, but since the secondary $\tau$ lepton
(unlike a secondary muon from a $\nu_\mu$ interaction) decays before it is
stopped in matter, new high-energy neutrinos can be produced from its decay
(``$\tau$ regeneration''). We use {\tt TAUOLA}~\cite{Jadach:1993hs} to simulate
$\tau$ decay, and propagate all secondary high-energy neutrinos out of the Sun
individually.

We compute the expected event rate in the IceCube detector by multiplying the
differential muon neutrino and antineutrino fluxes at the Earth by the effective
detector area $A_{\rm eff}(E_\nu)$~\cite{IceCube:2011rv} and then integrating over
energy. Note that we treat the neutrinos arriving at the Earth as a completely
incoherent mixture of mass eigenstates (see section~\ref{sec:nuosc}).
We have checked that neutrino absorption in the Earth (``Earth
shadowing''~\cite{Arguelles:2010yj,Koers:2008hv}) is negligible for our results.
We also do not need to consider oscillations in the Earth, which are
only relevant for neutrino energies close to one of the terrestrial MSW resonance.
However, the two standard resonances driven by $\Delta m_{21}^2$ and
$\Delta m_{31}^2$ are relevant only at neutrino energies below the IceCube energy
threshold, whereas the resonances mixing active neutrino and eV-scale sterile
neutrinos affect only $\mathcal{O}(\text{TeV})$ neutrinos, which cannot even
leave the Sun efficiently (see figure~\ref{fig:survival-prob}).
Note that the effective area given in~\cite{IceCube:2011rv} has been computed
from a simulation of the full 86-string IceCube detector, whereas the latest
published dark matter limits from IceCube are based on data taken in the
40-string IceCube configuration and in the older AMANDA-II detector. Since we
will ultimately use our simulation only to compute \emph{ratios} of event rates
between different oscillation models, we expect the systematic bias introduced
that way to be small. Note also that $A_{\rm eff}(E_\nu)$ as given
in~\cite{IceCube:2011rv} is the combined effective area for neutrinos and
antineutrinos. Since in sterile neutrino models, the relative importance of
neutrinos and antineutrinos in the IceCube signal changes, we need seperate
effective areas for neutrinos ($A_{\rm eff}^\nu(E_\nu)$) and antineutrinos ($A_{\rm
eff}^{\bar\nu}(E_\nu)$). We obtain them according to
\begin{align}
  A_{\rm eff}^\nu(E_\nu) = A_{\rm eff}(E_\nu)
    \frac{\sigma_{\rm CC}^\nu(E_\nu) \, d_{\mu^-}(E_\mu)}
         {\sigma_{\rm CC}^\nu(E_\nu) \, d_{\mu^-}(E_\mu)
        + \sigma_{\rm CC}^{\bar\nu}(E_\nu) \, d_{\mu^+}(E_\mu)} \,, \\
  A_{\rm eff}^{\bar\nu}(E_\nu) = A_{\rm eff}(E_\nu)
    \frac{\sigma_{\rm CC}^{\bar\nu}(E_\nu) \, d_{\mu^+}(E_\mu)}
         {\sigma_{\rm CC}^\nu(E_\nu) \, d_{\mu^-}(E_\mu)
        + \sigma_{\rm CC}^{\bar\nu}(E_\nu) \, d_{\mu^+}(E_\mu)} \,,
\end{align}
with the charged current neutrino--nucleon (antineutrino--nucleon) cross section
$\sigma_{\rm CC}^\nu(E_\nu)$ ($\sigma_{\rm CC}^{\bar\nu}(E_\nu)$), and the
muon (antimuon) range $d_{\mu^-}(E_\mu)$ ($d_{\mu^+}(E_\mu)$).
For simplicity, we assume a one-to-one relation between
the neutrino energy $E_\nu$ and the secondary muon energy $E_\mu$: $E_\mu =
(1 - y_{\rm CC}(E_\nu)) E_\nu$, where $y_{\rm CC}$ is the mean charged current
inelasticity parameter~\cite{Gandhi:1998ri}. We have checked that using full
differential cross sections would not significantly change our results.

We have verified our Monte Carlo code by comparing its predictions to published
results from~\cite{Cirelli:2005gh, Blennow:2007tw, WimpSimHP, Giunti:2009xz}.

While the advantage of the Monte Carlo technique is certainly its flexibility,
it is also quite computationally intensive.  Since we are also interested in
carrying out parameter scans over different sets of sterile neutrino parameters
(see section~\ref{sec:paramscans} below), we have also developed a faster code,
which does not take into account $\tau$ regeneration and energy loss in neutral
current interactions. Instead, it simply considers all neutrinos that interact
in the Sun in any way (NC or CC) to be lost to detection. Thus, for each given
set of oscillation parameters and for each neutrino energy, we need to solve
the equation of motion only once to determine the oscillation probabilities for
those neutrinos which do not interact. At each Runge-Kutta step, we also keep
track of the interaction probability to obtain simultaneously the fraction of
neutrinos at the considered energy which leave the Sun without interacting.

\begin{figure}
  \begin{center}
    \includegraphics[width=8cm,type=pdf,ext=.pdf,read=.pdf]{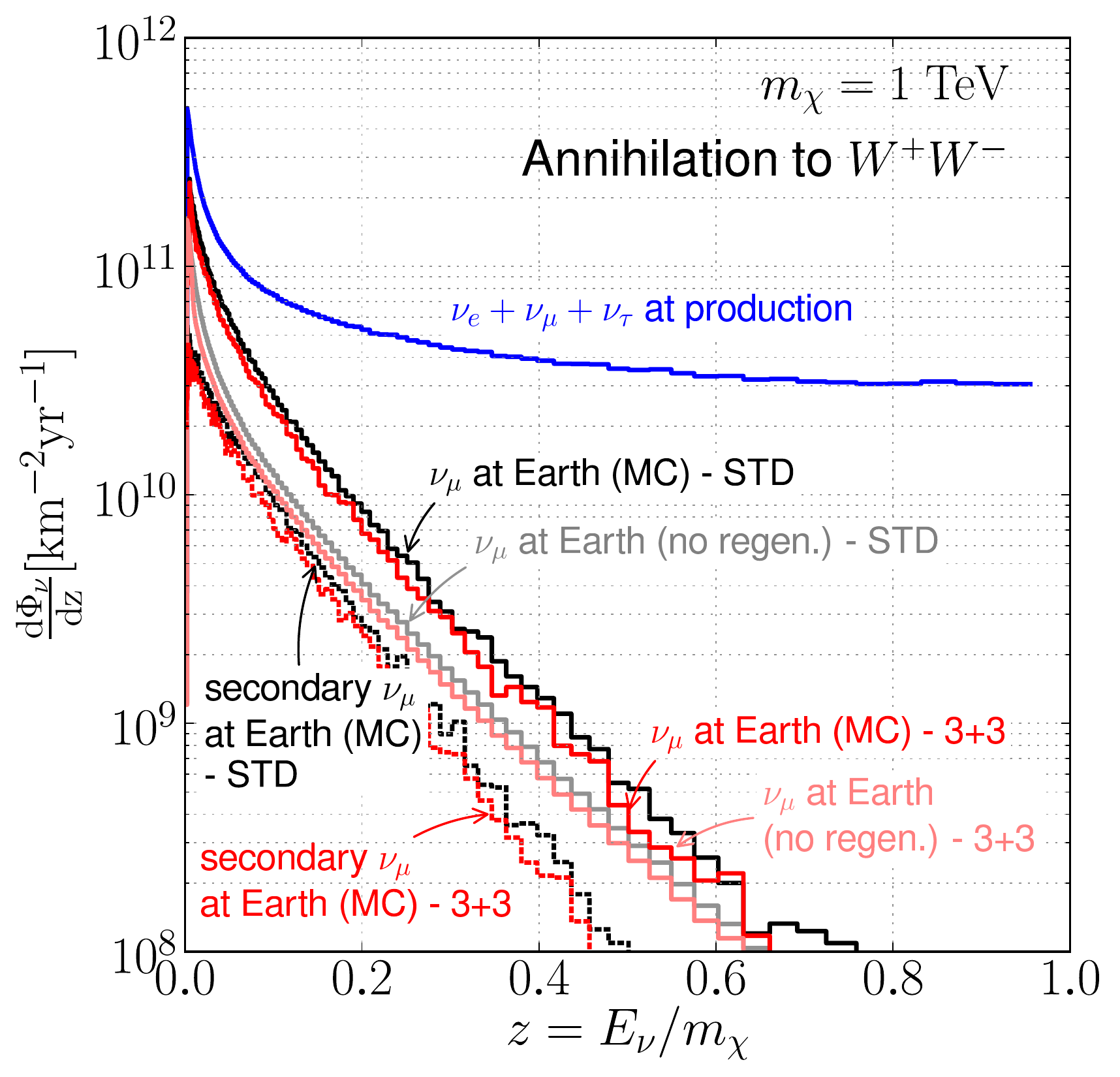}
    \includegraphics[width=8cm,type=pdf,ext=.pdf,read=.pdf]{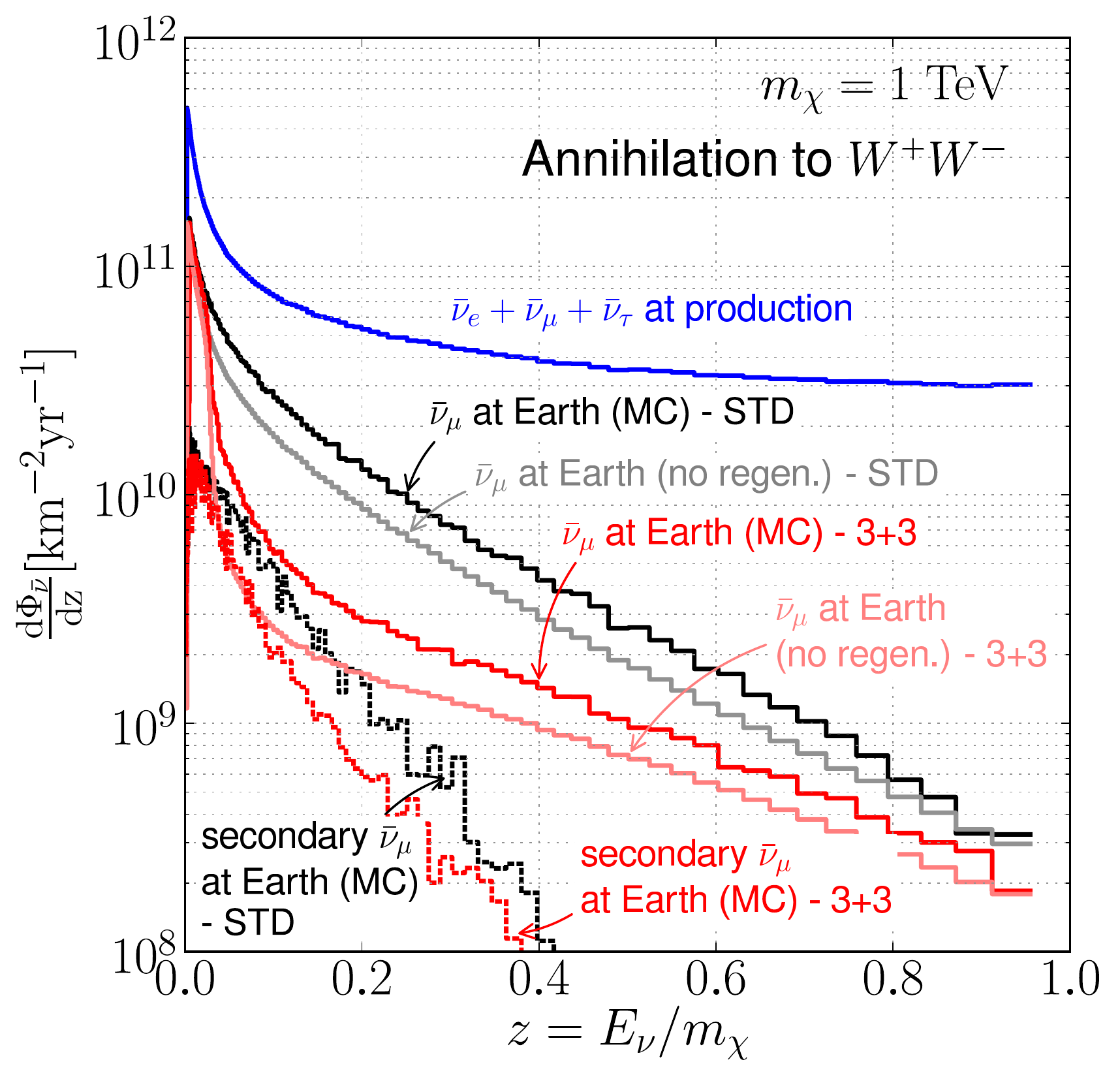}
  \end{center}
  \caption{Predicted neutrino fluxes (left) and antineutrino fluxes (right)
    for annihilation of a 1~TeV WIMP into $W^+ W^-$ in the Sun. We show the
    total neutrino flux at production, as well as the muon neutrino flux at the
    Earth. For illustration, we also show the flux of secondary neutrinos from
    $\tau$ regeneration, as well as the flux obtained using the simplified
    calculation that neglects regeneration and partial energy loss (see text
    for details). Results for standard oscillations (``STD'') are shown in
    black, results for the $3+3$ toy model introduced in
    section~\ref{sec:nuosc} are shown in red.}
  \label{fig:flux}
\end{figure}

We compare the results of our full Monte Carlo simulation to those of the
simplified method in figure~\ref{fig:flux}. We also show the flux of secondary
neutrinos from $\tau$ regeneration, and we notice that these neutrinos account
for most of the difference between the MC results and the ones from the
simplified method. (Another small contribution to this difference comes from
neutrinos that have undergone NC scattering, but are still within the
accessible energy range.) This conclusion is the same for standard three-flavor
oscillations (black and gray curves in figure~\ref{fig:flux}) and for
the $3+3$ model (red curves).

\section{Modified IceCube limits on dark matter capture in the Sun}
\label{sec:limits}

\begin{figure}
  \begin{center}
    \includegraphics[width=10cm]{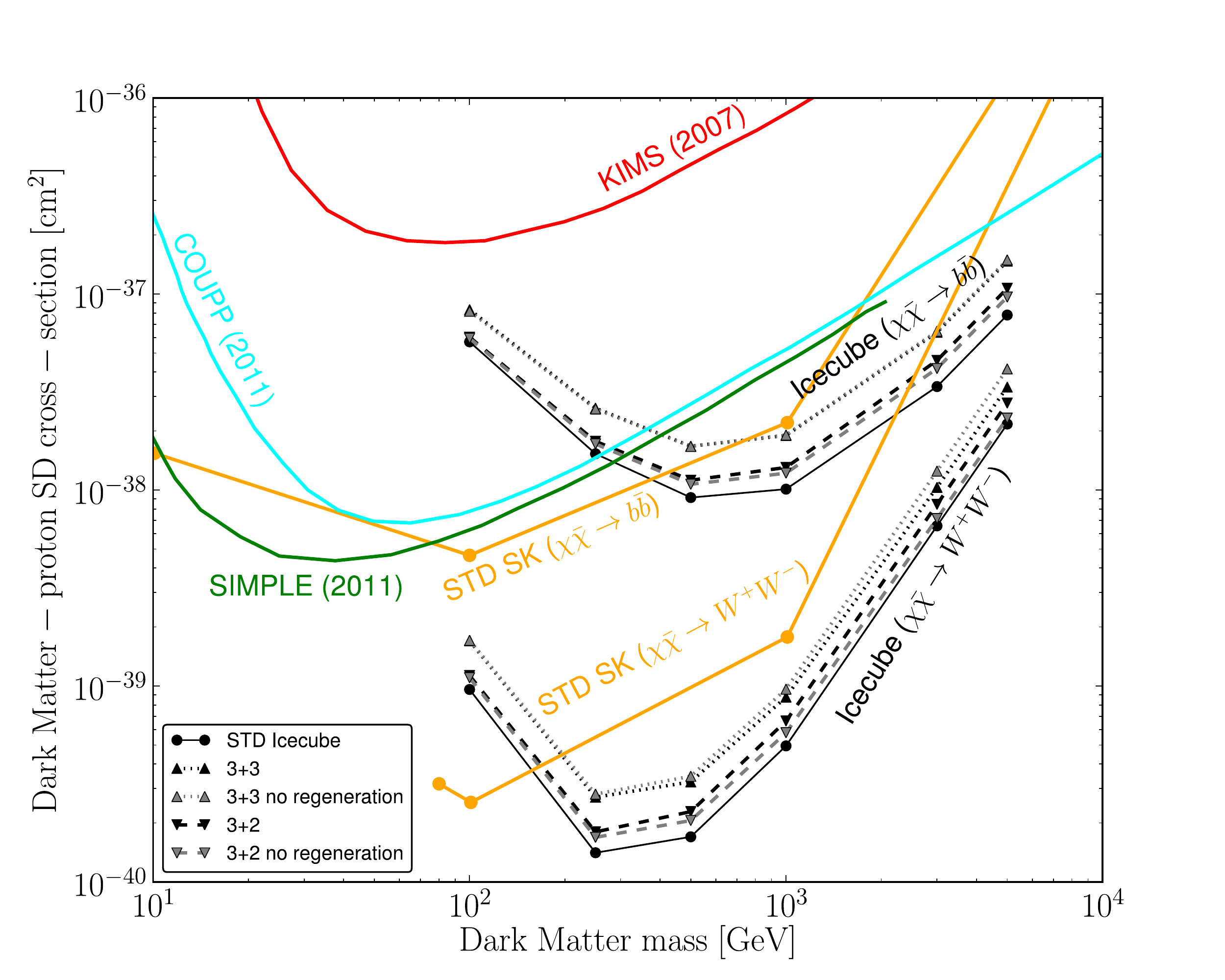}
  \end{center}
  \caption{IceCube limits on spin-dependent dark matter-proton
    scattering~\cite{IceCube:2011rv,IceCube:2011ec} in scenarios with sterile
    neutrinos (black/gray dashed and dotted lines) and in the standard
    oscillation (``STD'') framework (black/gray solid lines) compared to data
    from direct detection experiments~\cite{Lee:2007qn, Behnke:2010xt,
    Felizardo:2011uw, Desai:2004pq} and from Super-Kamiokande
    (SK)~\cite{Tanaka:2011uf} (colored lines).  Black lines correspond to
    results based on our Monte Carlo (MC) code, whereas gray lines are based on
    a simplified calculation which does not include secondary neutrinos (see
    text for details).  We see that for the $3+2$ scenario which provides the
    best fit to short baseline neutrino oscillation data, the limits are only
    moderately weakened. Our $3+3$ toy model, on the other hand, illustrates
    that larger modifications are possible.}
  \label{fig:IC-limit}
\end{figure}

In figure~\ref{fig:IC-limit} we show how the IceCube limits on spin-dependent
dark matter--proton scattering need to be modified if sterile neutrinos exist
(black and gray lines). For comparison we also show as colored lines limits
from a number of direct dark matter searches.  We have chosen the case of
spin-dependent dark matter scattering here rather than the more common
spin-independent interactions since the power of the IceCube limits compared to
direct searches is greater in this case~\cite{IceCube:2011ec}. We expect the
corrections to the IceCube limits due to sterile neutrinos to be very similar
in the two cases, though.

Solid black lines in figure~\ref{fig:IC-limit} are the published IceCube limits
from~\cite{IceCube:2011rv,IceCube:2011ec}; Dashed and dotted lines show the
constraint obtained in the $3+2$ model and the $3+3$ toy model introduced in
section~\ref{sec:nuosc}, respectively.  To obtain these results, we have used
the methods described in section~\ref{sec:sim} to predict the ratio of the
event rates at IceCube with and without sterile neutrinos, and we have then
rescaled the published IceCube 90\%~CL limit on the dark matter--nucleon
scattering cross section $\sigma_{90,\text{STD}}$ (which was computed assuming
standard oscillations) by this ratio. Specifically, if we denote the IceCube
event rate by $N_{\text{STD}}$, $N_{3+2}$ and $N_{3+3}$ for the standard
oscillation, $3+2$, and $3+3$ scenarios, respectively, we compute the cross
section limits in the $3+2$ and $3+3$ scenarios, $\sigma_{90,3+2}$ and
$\sigma_{90,3+3}$, according to
\begin{align}
  \sigma_{90,3+2} &= \sigma_{90,\text{STD}} \frac{N_{3+2}}{N_{\text{STD}}} \,, \\
  \sigma_{90,3+3} &= \sigma_{90,\text{STD}} \frac{N_{3+3}}{N_{\text{STD}}} \,.
  \label{eq:sigma-rescaling}
\end{align}
We see that the $3+2$ model leads to a moderate weakening of the cross section
limit, which can be understood from the fact that only electron neutrinos $\nu_e$
and muon antineutrinos $\bar\nu_\mu$ are substantially transformed into sterile
states (see figure~\ref{fig:prob-3+2}), and that these transitions happen only for neutrinos
with energies above several hundred GeV, whose contribution to the muon flux
at IceCube is suppressed due to the large absorption probability in the Sun.
In the $3+3$ toy model, on the other hand, resonant flavor transitions happen
already at lower energy (see figure~\ref{fig:prob-3+3}), and they happen for
$\nu_e$, $\bar\nu_\mu$ and $\bar\nu_\tau$.

As mentioned in section~\ref{sec:nuosc}, the effect could be even stronger if
$\Delta m_{41}^2$, $\Delta m_{51}^2$, and $\Delta m_{61}^2$ were negative
(which might, however, require non-standard cosmology to be consistent).

\section{Dependence on sterile neutrino parameters}
\label{sec:paramscans}

In section~\ref{sec:limits} we have illustrated using two exemplary models how
neutrino limits on dark matter capture and annihilation in the Sun are modified
by oscillations into sterile neutrinos. We are now going to study more
systematically how the worsening of these limits depends on the sterile
neutrino parameters. We do this using the $3+3$ toy model introduced in
section~\ref{sec:nuosc} since this model has only two new parameters
($\theta_s$ and $\Delta m_s^2$), but still covers the most important
phenomenological aspects of more general sterile neutrino scenarios.

\begin{figure}
  \begin{center}
    \includegraphics[width=12cm]{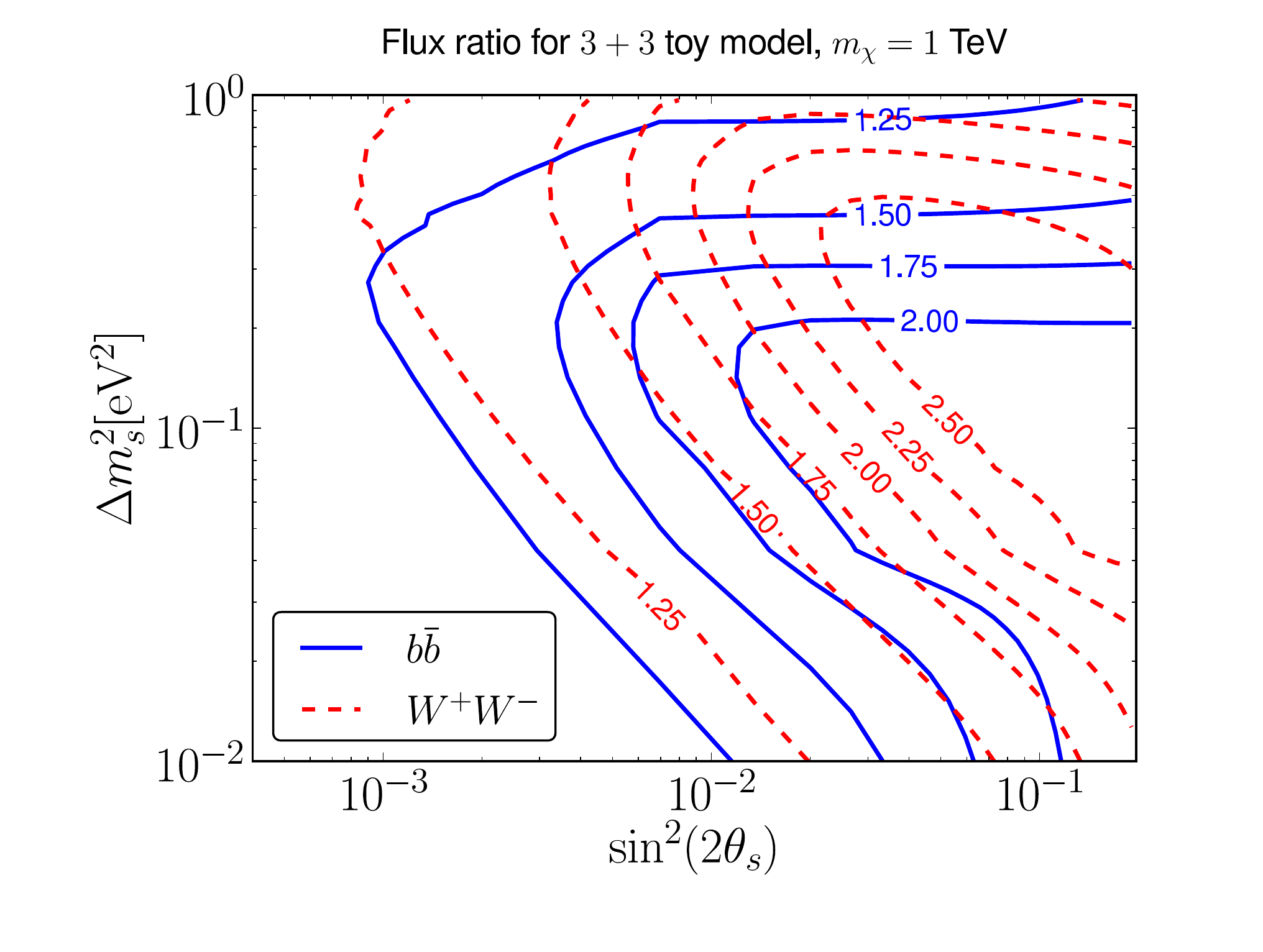}
  \end{center}
  \vspace{-0.5cm}
  \caption{Weakening of IceCube limits on dark matter capture and annihilation
    in the Sun due to sterile neutrinos, assuming for illustrative purposes the
    $3+3$ toy model introduced in section~\ref{sec:nuosc}. The contours show
    the factor by which the IceCube limit on spin-dependent dark matter--proton
    scattering cross section for a 1~TeV WIMP.  Red dashed contours are for
    annihilation into $W^+ W^-$ (which yields a rather hard neutrino spectrum),
    blue solid contours are for annihilation into $\bar{b} b$ (which yields a
    much softer spectrum). At large $\Delta m_s^2$, oscillations into sterile
    neutrinos become less relevant because the active--sterile MSW resonances
    move to very high energies; at small $\Delta m_s^2$ or small $\sin^2
    2\theta_s$, the MSW transitions become non-adiabatic.}
  \label{fig:3+3-paramscan}
\end{figure}

We show in figure~\ref{fig:3+3-paramscan} the factor by which the IceCube
limits on the spin-dependent dark matter--nucleon scattering cross section are
weakened for a wide range of $\sin^2 2\theta_s$ and $\Delta m_s^2$ values. The
shape of the contours can be understood as follows: At very large $\Delta
m_s^2$, the new MSW resonances,
equations~\eqref{eq:MSW-3+3-e}--\eqref{eq:MSW-3+3-tau} lie at a very high
neutrino energy. For instance, at $\Delta m_s^2 = 1$~eV$^2$,
equation~\eqref{eq:MSW-3+3-e} yields a resonance energy of about 60~GeV at
solar core densities, i.e.\ only neutrinos with $E_\nu \gtrsim 60$~GeV are affected
by the resonance.  Since very high energy neutrinos are mostly absorbed in the
Sun, they do not contribute significantly to the IceCube limits. For somewhat
lower $\Delta m_s^2$, the resonances move down in energy into the region
relevant to IceCube. For too low $\Delta m_s^2$ or for too small $\theta_s$, on
the other hand, MSW-enhanced flavor transitions become non-adiabatic (see
equation~\eqref{eq:adiabaticity} and related discussion), suppressing
active--sterile transitions again. This happens first at high energy, which is
why at low $\Delta m_s^2$ the correction factors shown in
figure~\ref{fig:3+3-paramscan} are generally larger for dark matter
annihilation into the soft $\bar{b} b$ channel than for annihilation in to the
hard $W^+W^-$ final state.

\section{Discussion and conclusions}
\label{sec:conclusions}

In this paper, we have shown how IceCube limits on dark matter capture and
annihilation in the Sun are modified if eV-scale sterile neutrinos exist, as
suggested by part of the short baseline oscillation data. Since IceCube is
looking for high-energy neutrinos from dark matter annihilation in the center
of the Sun, its results depend strongly on the oscillations of these neutrinos
on their way out of the Sun. We have argued that in sterile neutrino scenarios
new high-energy MSW resonances can lead to almost complete conversion of
certain neutrino flavors into sterile states inside the Sun.  In this case
IceCube's constraints on dark matter--nucleon scattering can be significantly
weakened, by a factor of two or more.

This may have interesting implications if in the future dark matter is detected
in a direct search or at the LHC, but the parameters determined there are in
conflict with limits (or signals) from neutrino telescopes.  If the allowed
dark matter annihilation channels and branching fractions are established at
the LHC, such a conflict could then provide a clear and strong hint for the
existence of sterile neutrinos. With sufficient data, neutrino telescopes would
even be able to contribute the determination of the active--sterile mixing
parameters.

\emph{Note added:} While we were completing this work,
reference~\cite{Esmaili:2012ut} appeared on the arXiv, addressing similar
topics.

\begin{acknowledgments}
It is a pleasure to thank Matthias Danninger for very useful discussions on the
IceCube dark matter search.  We are also grateful to the Direcci\'{o}n de
Inform\'{a}tica Acad\'{e}mica at the Pontificia Universidad Cat\'{o}lica del
Per\'{u} (PUCP) for providing distributed computing support through the LEGION
system.  CA would like to thank Fermilab for warm hospitality and for support
through the Latin American Students Program during his six month visit in
summer 2011.  JK is grateful to the Aspen Center for Physics (supported by the
National Science Foundation under Grant No.~1066293), where part of this work
has been carried out.  CA is supported by the Direcci\'{o}n de Gesti\'{o}n de
la Investigaci\'{o}n at PUCP through grant DGI-2011-0180.  Fermilab is operated
by Fermi Research Alliance, LLC, under contract DE-AC02-07CH11359 with the
United States Department of Energy.
\end{acknowledgments}

\appendix
\section{Numerics of neutrino oscillation probabilities}
\label{sec:numerics}

In this appendix, we discuss the algorithm used to compute the neutrino
oscillation probabilities in the Sun. As mentioned in section~\ref{sec:sim} we
use an implementation of the well known Runge-Kutta (RK) algorithm~\cite{NUM},
namely the {\tt rkf45} algorithm implemented in the GNU Scientific
Library~\cite{Galassi:GSL}. In each iteration this algorithm uses a step
function to evolve the neutrino state vector from a time $t_0$  to a time
$t_0+\Delta t$ by approximately solving the Schr\"{o}dinger equation, where
$\Delta t$ is chosen such that the optimal balance between speed and accuracy
is achieved. Rather than working entirely in one basis, we transform the
Schr\"{o}dinger equation to an instantaneous interaction basis before each
step. This instantaneous interaction basis is defined by the transformation 
\begin{align}
  \psi_I(t;t_0) = S(t,t_0) \, \psi(t) \equiv e^{iH_0 (t-t_0)} \psi(t) \,,
\end{align}
where the Hamiltonian has been separated in the following manner
\begin{align}
  H(t) = H(t_0)+\Delta H(t;t_0)\,,
\end{align}
with 
\begin{align}
  H_0(t_0) = \frac{1}{2E_\nu}U D U^\dagger + V(t_0) \,,\qquad
  \Delta H(t;t_0) = V(t) - V(t_0) \,.
\end{align}
Here $V(t)$ is the neutrino matter potential (see equation~\eqref{eq:MSW}), $E_\nu$
the neutrino energy, $U$ the leptonic mixing matrix, and $D=\diag(0,\Delta
m^2_{21},\Delta m^2_{31},...)$.

The Schr\"{o}dinger equation in the interaction basis is
\begin{align}
  i \frac{d\psi_I}{dt} = H_I \psi_I
\end{align}
with $H_I(t;t_0) =S(t,t_0) \, \Delta H \, S^\dagger (t,t_0) $.  Since the
matter potential changes slowly in the Sun and thus $H_I$ is small, the RK
algorithm can choose a larger step size $\Delta t$ compared to a calculation in
the flavor basis.  Note that the elements $S_{jk}$ of the transformation matrix
$S(t,t_0)$ can be evaluated efficiently by computing $\tilde{V}_{jm} e^{-i
\lambda_m (t-t_0)} (\tilde{V}^\dagger)_{mk}$, where $\lambda_m$ are the eigenvalues of
$H_0$, and $\tilde{V}$ is the matrix of the corresponding eigenvectors.  After the
evolution of the step has concluded we transform $\psi_I$ back to the flavor
basis,
\begin{align}
  \psi(t_0+\Delta t) = e^{-iH_0(t_0) \, \Delta t} \psi_I(t_0+\Delta t) \,,
\end{align}
and proceed to the next step, setting $t_0 \to t_0 + \Delta t$.

\bibliographystyle{apsrev4-1}
\bibliography{wimp-ann}

\end{document}